\documentclass[useAMS,usenatbib]{mn2e}
\usepackage{graphicx}

\newcommand{\dex}[1]{\mbox{$\times 10^{#1}$}}

\newcommand{\cntss}{counts s$^{-1}$}

\newcommand{\nbox}{\mbox{$n_{\mbox{\footnotesize box}}$}}

\newcommand{\nspike}{\mbox{$n_{\mbox{\footnotesize spk}}$}}

\newcommand{\ebv}{\mbox{$E\left(\mbox{B$-$V}\right)$}}

\newcommand{\mab}{\mbox{$m_{\mbox{\footnotesize AB}}$}}
\newcommand{\logg}{\mbox{$\log{\left(g\right)}$}}

\newcommand{\galex}{{\it GALEX}}
\newcommand{\om}{{XMM-OM}}

\title[XMM-Newton serendipitous UV source survey]{The XMM-Newton
  serendipitous ultraviolet source survey catalogue}

\author[M.J. Page et al.]
{M.J. Page$^{1}$, C. Brindle$^{1}$, A. Talavera$^{2}$, 
M. Still$^{1,3}$, S.R. Rosen$^{1,4}$, 
V.N. Yershov$^{1}$, 
\and H. Ziaeepour$^{1,5}$, K.O. Mason$^{1,6}$, M.S. Cropper$^{1}$, 
A.A. Breeveld$^{1}$, N. Loiseau$^{2}$, 
\and R. Mignani$^{1,7}$, A. Smith$^{1}$,  P. Murdin$^{8}$\\
$^{1}$Mullard Space Science Laboratory, University College London,
Holmbury St Mary, Dorking, Surrey, RH5 6NT, UK\\
$^{2}$XMM-Newton Science Operations Centre, ESA, Villafranca del Castillo, 
Apartado 78, 28691 Villanueva de la Ca\~{n}ada, Spain\\
$^{3}$NASA Ames Research Center, Moffett Field, CA 94035, USA\\
$^{4}$Department of Physics and Astronomy, University of Leicester,
Leicester LE1 7RH, UK\\
$^{5}$Max Planck Institut f\"ur Extraterrestrische Physik (MPE), 
Giessenbachstrasse 1, 85748 Garching, Germany\\
$^{6}$Science and Technology Facilicities Council, Polaris House, 
North Star Avenue, Swindon, Wilts SN2 1SZ, UK\\
$^{7}$Kepler Institute of Astronomy, University of Zielona G\'ora, Lubuska 2, 
65-265 Zielona G\'ora, Poland\\
$^{8}$Institute of Astronomy, Madingley Road, Cambridge CB3 0HA, UK
}

\begin{document}

\date{Accepted ----. Received ----; in original form ----}

\pagerange{\pageref{firstpage}--\pageref{lastpage}} 
\pubyear{2011}
\maketitle

\label{firstpage}

\begin{abstract}
  The {\em XMM-Newton} Serendipitous Ultraviolet Source Survey (XMM-SUSS) is
  a catalogue of ultraviolet (UV) sources detected serendipitously by
  the Optical Monitor (XMM-OM) on-board the {\it XMM-Newton}
  observatory. The catalogue contains ultraviolet-detected sources
  collected from 2\,417 XMM-OM observations in 1--6 broad band UV and
  optical filters, made between 24 February 2000 and 29 March 2007.
  The primary contents of the catalogue are source positions,
  magnitudes and fluxes in 1 to 6 passbands, and these are
  accompanied by profile diagnostics and variability statistics. The
  XMM-SUSS is populated by 753\,578 UV source detections above a
  3$\sigma$ signal-to-noise threshold limit which relate to 624\,049
  unique objects.  Taking account of substantial overlaps between
  observations, the net sky area covered is 29--54 deg$^2$, depending
  on UV filter.  The magnitude distributions peak at \mab\ = 20.2,
  20.9 and 21.2 in 
  UVW2 ($\lambda_{eff}=2120$\AA), UVM2
  ($\lambda_{eff}=2310$\AA) and UVW1 ($\lambda_{eff}=2910$\AA)
  respectively. More than 10 per cent of sources have been visited
  more than once using the same filter during {\em XMM-Newton} 
  operation, and $>
  20$ per cent of sources are observed more than once per filter
  during an individual visit. Consequently, the scope for science
  based on temporal source variability on timescales of hours to years
  is broad. By comparison with other astrophysical catalogues we test
  the accuracy of the source measurements and define the nature of the
  serendipitous UV XMM-OM source sample. The distributions of source
  colours in the UV and optical filters are shown together with the
  expected loci of stars and galaxies, and indicate that sources which
  are detected in multiple UV bands are predominantly star-forming
  galaxies and stars of type G or earlier.

\end{abstract}


\section{Introduction}
\label{sec:introduction}

One of the instruments carried by the European Space Agency's {\it
  XMM-Newton} satellite is the Optical Monitor, an ultraviolet/optical
telescope with a 30cm diameter primary mirror \citep{mason01}. 
While its main rationale is to provide complementary data to
those from the X-ray instruments, and in particular to create a
simultaneous multi-wavelength capability to constrain spectral energy
distributions, {\it XMM-Newton} Optical Monitor (XMM-OM) is a 
capable
instrument in its own right. With a field of 17x17 arcmin$^{2}$ and a
full width half maximum of the point spread function of less than 2
arcsec 
over the full field of view, 
XMM-OM provides a powerful UV survey 
capability over much
larger fields than is possible with the {\it Hubble Space Telescope}
UV instrumentation, and at a finer spatial sampling than provided by
the {\it GALEX} satellite \citep{martin05}.

A UV survey capability for XMM-OM was envisaged at the inception of
the instrument concept in 1988, and the UV filter choice and
observation strategies were planned to maximise the UV survey science
return. At that time, in addition to sounding rocket observations, UV
photometric surveys had been carried out by the {\it Orbiting
Astronomical Observatory 2} \citep{code70,davis72},
{\it TD-1} \citep{boksenberg73,dejager74} and 
the {\it Astronomical Netherlands Satellite} \citep{vanduinen75}. 
They were augmented by other surveys
including from {\it Skylab} \citep{henize75}, {\it Apollo 
16} \citep{carruthers73} and {\it Apollo 17} \citep{henry75}. These
had provided first generation views of the UV sky. While the {\it 
IUE} satellite launched in 1978 \citep{boggess78}
carried out more than $10^5$ UV spectroscopic observations over a
period of 17 years, 
the imaging, and therefore large-scale knowledge of the UV sky remained
surprisingly limited.

During the 1990s, a broader but still limited view of the UV sky was
achieved using imagers on the Space Shuttle (FAUST, Bowyer {\it
  et~al.,} 1993, UIT, Stecher {\it et al.,} 1997), and then much more
detailed views using {\it HST}, initially with the Faint Object Camera
\citep{albrecht02}, 
followed by the Wide Field and Planetary Camera 2
\citep{trauger94}, the Space Telescope Imaging Spectrograph
\citep{kimble98}, the Advanced Camera for Surveys \citep{sirianni05}
and the Wide Field Camera 3 \citep{mackenty10}.

{\it XMM-Newton} with XMM-OM was launched in 1999, 
  followed in 2003 by {\it GALEX} \citep{martin05} and in 2005 by 
  {\it Swift} with its Ultraviolet and Optical Telescope
  \citep[UVOT]{roming05}, of a similar design to XMM-OM. The imaging
  passbands of XMM-OM and {\em GALEX} are given in
  Table~\ref{tab:filters}; the UVOT on {\em Swift} has similar
  passbands to XMM-OM.
The era of large-scale
UV surveys to faint limiting magnitudes had arrived. {\it GALEX} in
particular has sampled nearly the full sky, outside of the Galactic
plane, through two broadband filters covering the near-UV and far-UV
respectively.

This paper describes the {\it XMM-Newton} Serendipitous Ultraviolet
Source Survey (XMM-SUSS), an electronic catalogue of UV sources
detected serendipitously by the XMM-OM. 
The catalogue contains 
ultraviolet source
detections collected from 2\,417 {\it XMM-Newton} observations made
between 24 February 2000 and 29 March 2007. Taking account of
substantial overlaps between observations, the net sky area covered ranges from
29 to 54 deg$^2$ depending on UV filter. 
The primary content of the
catalogue is source positions and photometry in up to 3
UV filters and 3 optical filters for 624\,049 unique UV sources. 
This is accompanied by spatial extent and variability
diagnostics.

With respect to the {\it GALEX} All-sky Imaging Survey (AIS) and
Medium Imaging Survey (MIS), the XMM-SUSS covers a much smaller sky
area, reaching magnitudes intermediate between the AIS and MIS, but
has finer photometric sampling in wavelength, better morphological
discrimination, and less susceptibility to source confusion. The
latter point, which is a consequence of the XMM-OM having a much finer
instrumental point spread function than {\it GALEX} (FWHM $\le 2$
arcsec compared to 4--5 arcsec; see Table \ref{tab:filters}), is
particularly important for the fidelity of the XMM-SUSS in crowded
regions such as the Galactic plane and Magellanic Clouds.  The
XMM-SUSS also has the advantage that the UV observations were obtained
simulaneously with sensitive X-ray imaging, the source content of
which is easily accessible through the 2XMM catalogue
\citep{watson09}.  Furthermore, many of the XMM-SUSS sources were
observed more than once per filter during an {\it XMM-Newton}
observation and/or through multiple {\it XMM-Newton}
visits. Consequently, the scope for science based on temporal source
variability on timescales of hours to years is broad.

XMM-SUSS is a release originating from the XMM-OM instrument team on
behalf of ESA and coordinated with the {\it XMM-Newton} Science
Survey Centre \citep[SSC, ][]{watson01}. An alternative, independent 
catalogue of XMM-OM sources (OMCat), based on an earlier version of the 
{\it XMM-Newton} Science Analysis Software (SAS), is described in
\citet{kuntz08}. 
Since 2008, 
the XMM-SUSS has been available from the 
{\it XMM-Newton} Science Archive\footnote{http://xmm.esac.esa.int/xsa/},
NASA HEASARC\footnote{http://heasarc.gsfc.nasa.gov}, the XMM-SUSS project 
pages\footnote{http://www.mssl.ucl.ac.uk/www\_astro/XMM-OM-SUSS} 
and through the VO
interfaces.  An initial announcement of the catalogue was made 
in \citet{still08}.

This paper is laid out as follows. The instrumentation and data
selection are described in Section \ref{sec:sample} and the data
processing is described in Section \ref{sec:reduction}. In Section
\ref{sec:validation} we describe the tests which were carried out to
check the quality and reliability of the catalogue. The properties of
the catalogue are described in Section \ref{sec:content}. A brief
description of the differences between the XMM-SUSS and the OMCat is
provided in Section \ref{sec:omcat}. Known issues wih the XMM-SUSS, and our plans for future development are described in Section \ref{sec:knownissues}. Our conclusions are presented in
Section \ref{sec:conclusions}. Testing and validation of the source 
detection algorithm used for XMM-SUSS are given in Appendix A. 
Details of the quality flagging
algorithms are given in Appendix B and a list and description of the
columns in the XMM-SUSS source table are given in Appendix C.

\section{Instrumentation and data sample}
\label{sec:sample}

\subsection{The XMM-Newton Optical Monitor}
\label{subsec:om}

The XMM-OM is a 30-cm  optical/UV   
telescope  of  a   modified  Ritchey  Chr\'{e}tien
design, coaligned with the X-ray telescopes \citep{mason01}. 
From  the primary  mirror, incoming light  is reflected  via the
secondary mirror onto  a rotatable, flat mirror which directs the
beam onto one of two identical filter wheel and detector assemblies.

Each of the \om\ detectors 
consists 
of a photon-counting micro-channel
plate (MCP) intensified Charge Coupled Device (CCD) \citep{fordham89}.
Incident photons eject electrons from a multi-alkali (S20)
photo-cathode, optimized to UV and blue wavelengths, which are
amplified by a factor $10^6$ in number using MCPs in series. These
electrons strike a phosphor screen, the photons from which are fed to
the CCD via a fibre-optic taper. By centroiding the photon cascade on
the CCD, the sky positions of incident photons are determined by
on-board software to a degree of precision which is much higher than
the physical CCD pixel size.  An image is constructed in real time 
on-board which
subsamples the CCD array by factor 8, providing data of $2048 \times
2048$ pixels of size $0.476 \times
0.476$~arcsec$^{2}$.  The onboard centroiding gives rise to a 
low-level modulo-8
fixed-pattern distortion \citep{kawakami94} which is routinely 
corrected in the ground processing. The detector response is linear
when the arrival rate of events per photon-cascade resolution 
element\footnote{the group of physical CCD pixels containing a photon splash, 
as distinct from the image resolution} is significantly below the
CCD readout frame rate, typically 90--140 s$^{-1}$. At higher count rates, 
the detector
response is non-linear, an effect known as coincidence-loss
\citep{fordham00}.  
This effect becomes significant ($\sim$~10 per cent) at around 0.1 counts 
per frame (typically 
corresponding to sources of 2.3 -- 2.8 magnitudes brighter than the zeropoints 
listed in Table \ref{tab:filters}).
It is corrected for during ground
processing, but as sources approach saturation 
(approximately 5.5 -- 6 magnitudes brighter than the zeropoints listed 
in Table \ref{tab:filters})
the
photometric measurement errors become larger, rather than smaller, 
with count rate \citep{kuin08}. At very high count rates, 
centroiding of multiple, overlapping photon splashes in each frame causes the point spread 
function to become distorted, and bright
sources are surrounded by regions of coincidence-loss-induced modulo-8 noise
which cannot be corrected in ground processing. 

The detector design results in zero readout noise and a low level of
dark noise.  Above the Earth's atmosphere the small aperture of the XMM-OM
is compensated by its UV sensitivity and the absence of atmospheric
extinction and diffraction.  Sky background is dominated by diffuse
zodiacal and Galactic light.  \om\ is therefore well-suited to detecting faint
sources. CCD pixel-to-pixel sensitivity variations are unimportant,
cosmetic CCD defects 
cause few image defects,
 and large scale
sensitivity gradients are small. Cosmic rays are discriminated and
eliminated on board.

\begin{table*}
 \centering
 \caption{Characteristics of the XMM-OM imaging passbands. The
   corresponding numbers for the {\it GALEX} FUV and NUV passbands are
   given for comparison; 
  {\it Swift UVOT} has similar passbands to XMM-OM.
   The wavelength ranges are for an effective
   area $> 10$ per cent of peak. Note that U filter has a small
   additional region of sensitivity around 4760\AA\ in addition to the
   wavelength range given below (see Fig.~\ref{fig:effarea}). 
   Fluxes in the XMM-SUSS correspond to the effective wavelengths listed.
   The corresponding numbers for {\em GALEX} are taken from 
   \citet{morrissey07}.
}
\label{tab:filters}
  \begin{tabular}{@{}lcccccccc@{}}
    \hline
    & \multicolumn{6}{c}{\it XMM-OM} & \multicolumn{2}{c}{\it GALEX}\\
    & UVW2 & UVM2 & UVW1 & U & B & V & FUV & NUV\\ 
    \hline
    Effective Wavelength (\AA) & 2120 & 2310 & 2910 & 3440 & 4500 & 5430 & 1539 & 2316 \\
    Wavelength range (\AA)
    & 1800--2550 & 1990--2700 & 2430--3610 & 3030--3890 & 3810--4900 & 5020-5870 & 
    1344--1786 & 1771--2831 \\
    Zero point (AB) & 16.57 & 17.41 & 18.57 & 19.19 & 19.08 & 17.92 & 18.82 & 20.08 \\
    Peak effective area (cm$^2$) & 3 & 8 & 20 & 42 & 39 & 23 & 37 & 62 \\
    FWHM resolution (arcsec) & 1.98 & 1.80 & 2.00 & 1.55 & 1.39 & 1.35 & 4.2 & 5.3 \\
    \hline
\end{tabular}
\end{table*}

Both the position
and arrival time of photons are recorded on-board. 
Non-dispersive
observations can be performed in  two modes, IMAGING mode (with no recorded
photon arrival times) and FAST mode, which time-tags each photon.  
IMAGING mode provides the largest
field-of-view available at the expense of timing
information. The largest image available is 17$\times$17 arcmin$^2$,
although  commonly smaller
windows are used  to either tailor an observation  to specific science
goals or to meet  telemetry or onboard storage limits. 
The allowed exposure times of IMAGING mode exposures range from 800~s 
to 5000~s. 
FAST mode
can be achieved within the onboard memory budget only if the window is
small, typically  $10.5 \times 10.5$~arcsec$^{2}$.
IMAGING mode data are typically obtained in parallel  during FAST mode
observing. The majority of IMAGING 
data are binned 
$2\times 2$ onboard to $1^{\prime\prime}\times1^{\prime\prime}$ image pixels.
In most cases the full
field of view is  sampled during an observation  in at least  one filter, either
using  a  full-frame IMAGING  mode  or  by  mosaicing smaller  IMAGING
windows obtained in series. Modes are discusseed more fully 
in \citet{mason01} and
\citet{ehle08}.

{\em XMM-Newton}'s 48h orbit allows for long, uninterrupted pointings of
science targets. The XMM-OM typically takes multiple exposures of the
same field through several filters, sometimes with a sequence of differing IMAGING mode windows
in order to cover the full field of view. Individual sources are often recorded more
than once through the same filter where either sub-windows overlap or
exposures are repeated.

\begin{figure}
\includegraphics[width=63mm,angle=270]{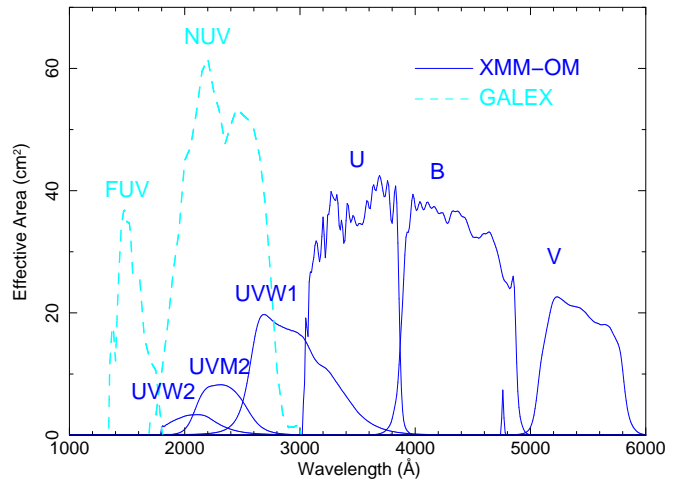}
\caption{Effective areas of the XMM-OM pass bands
  (solid-lines) together with those of {\it GALEX} (dashed-lines). The small peak at 
4760\AA\ is part of the U filter response. The sharp cutoff in the UVW2 pass band 
is due to the materials used in the optical elements rather than the filter itself.}
\label{fig:effarea}
\end{figure}

There are seven imaging filters mounted in the \om\ filter wheel
together with two grisms for low-dispersion spectroscopy.  One of the
filters, WHITE, transmits over the full XMM-OM bandpass
(1800--8000\AA) to maximise throughput. The remaining filters in the
order of increasing central wavelength are called UVW2, UVM2, UVW1, U,
B and V, where the final three filters cover similar wavelength ranges
to the Johnson UBV set \citep{johnson51}. The effective areas of the
XMM-OM imaging passbands are shown in Fig.~\ref{fig:effarea} together
with those of \galex. Although not visible in Fig.~\ref{fig:effarea},
the response curves of the UVW2 and UVM2 filters extend beyond
3000\AA\ into the optical range, and the UVW1 response curve extends
beyond 4100\AA, albeit with very low throughput ($<0.1$ cm$^{2}$). The
basic properties of the XMM-OM passbands are listed in Table
\ref{tab:filters}.

\subsection{Data selection}
\label{subsec:selection}

The XMM-SUSS is derived from 2\,417 {\it XMM-Newton} observations
obtained between 24 February 2000 and 29 March 
2007, all of which are publicly available in the 
{\it XMM-Newton} science archive. Because the XMM-SUSS is
a catalogue of UV sources, only those {\it XMM-Newton} observations which
include at least one XMM-OM exposure through the UVW2, UVM2 or UVW1
filters were considered for the catalogue.

The catalogue is constructed entirely from data taken in IMAGING mode. 
FAST mode data are excluded because of the small window
size. Similarly, $2\times2$ arcmin$^{2}$ unbinned central windows of the default
imaging configuration are ignored, although
note that in the majority of observations, the central region is also
recorded as an $8\times8$ arcmin$^{2}$ image which is included within
the catalogue sample. Data taken with the WHITE filter
provide very limited colour information and so are not utilised 
for the catalogue. Grism
data are not used for the catalogue because of the different nature of the
dispersed data and to avoid the source confusion within the grism
fields. 

The pointing of {\it XMM-Newton} is usually good to 3 arcsec. 
Boresight positions of each field are corrected by comparison
of XMM-OM sources with the USNO-B1.0 catalogue, although a minimum
number of source detections are required to perform the correction
accurately. Images were not used for the catalogue if they contain $<
5$ source detections, if the RMS residual between matched source
positions and their USNO-B counterparts is $> 1.5$ arcsec, or if the
$1\sigma$ error in the computed boresight correction is $> 1.0$ arcsec
in either the right ascension or declination directions.

Both the source detection algorithm and aperture photometry become
increasingly uncertain in crowded fields owing to source confusion and
the lack of background measures. Images were excluded from the
catalogue if they had a detected source density $>$ 35 arcmin$^{-2}$,
corresponding to 10\,000 sources within a full-frame image. 
The
majority of fields excluded by this criterion are U, B and V
observations of the Galactic plane and Magellanic Clouds.

\begin{figure}
\includegraphics[width=85mm,angle=0]{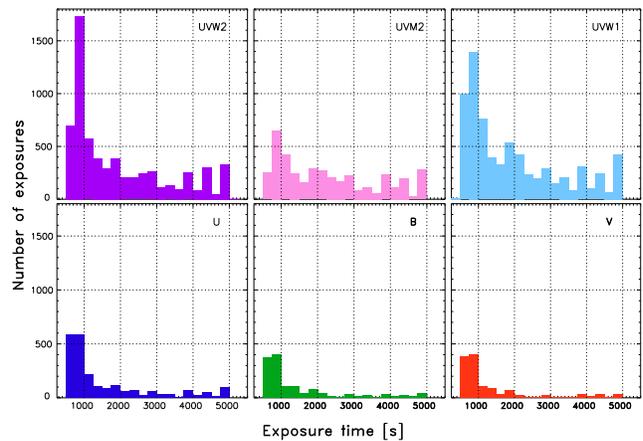}
\caption{Exposure times of the XMM-OM images used to construct the XMM-SUSS, 
split by filter.
}
\label{fig:exposuretimes}
\end{figure}

Finally, mosaiced images for each field were inspected visually for
obvious problems.  Images which were dominated by large scattered
light features from bright, off-axis sources, those with evidence of
telemetry corruption, and those which showed signs of spurious
attitude drift were excluded from the catalogue. 
The distribution of exposure times of the images which were used 
for the XMM-SUSS, 
by bandpass, are shown in Fig. \ref{fig:exposuretimes}.

\section{Data processing}
\label{sec:reduction}

The data were processed using software tasks from the {\em XMM-Newton}
SAS version 8.0.  
A number of improvements to the SAS, in particular to the source
detection task {\sc omdetect}, were driven by the development of the
XMM-SUSS and are implemented in the public release of the SAS
(versions 8.0 and higher) so that source lists of comparable quality
to the XMM-SUSS are generated by the standard {\em XMM-Newton}
pipeline and distributed as pipeline products, or can be produced by
members of the scientific community as required. Of the procedures
described in this section, only the merging and concatenation of the
source lists for the catalogue (Section \ref{subsec:merging}) is
carried out outside the standard suite of SAS tasks.

\subsection{Raw data}
\label{subsec:RawData}

A description of the raw XMM-OM data delivered by the {\em XMM-Newton}
spacecraft is available in \citet{guainazzi04}. Spacecraft pointing
history from independent on-board star trackers is telemetered down
with the XMM-OM data.
The XMM-OM also records tracking history data
from its own dedicated tracking mode windows.

Data required to convert the brightness, position and extent of
detected XMM-OM sources from detector units to physical units have been
calibrated by the XMM-OM Instrument Team and are stored in the 
{\it XMM-Newton} Current
Calibration File, which is accessible from the 
{\it XMM-Newton} Science Operations Centre web 
pages\footnote{http://xmm.esac.esa.int/}. The current state of the XMM-OM 
calibration is given by \citet{talavera11}.
 All raw data employed to construct the
XMM-SUSS catalog are in the public domain and can be downloaded from
the {\it XMM-Newton} Science Archive.

\subsection{Bad pixels}
\label{subsec:BadPixels}

No attempt is made to correct or interpolate over damaged or
contaminated detector pixels. Positions of pixels which do not image 
the sky (at the corners of the array) and detector defects (Fig.~\ref{fig:Mod8badpix}) 
are instead recorded 
within the Current
Calibration File. Bad pixels are masked from photometric measures, and sources
including bad pixels are flagged as
having additional unquantified uncertainties associated with their
brightness and location. It is left to the catalog user's discretion
whether to include or exclude such sources from their data samples.

\subsection{Bias, dark-current and spatial response variations}
\label{subsec:FlatField}

Owing to the photon-counting nature of the detector, there is no CCD
bias level to subtract from images. The dark count rate is below 
$5 \dex{-4}$ counts\,s$^{-1}$\,pix$^{-1}$ and varies across
the detector by $<10$ per cent. Therefore dark counts make only a small 
contribution
to the background compared to the zodiacal light, and no attempt is
made to separate these two background components within the data
processing. Source and background extraction regions are small and
from the same image locale, and hence the variation of the dark
current over the detector is not a significant factor for source extraction. To
within measurement uncertainties, the sensitivity is uniform 
(to within 5 per cent) 
with spatial position over the detector, 
and hence no correction is applied for spatial sensitivity variations.

\subsection{Fixed-pattern image structure}
\label{subsec:Mod8}

\begin{figure}
\includegraphics[width=84mm]{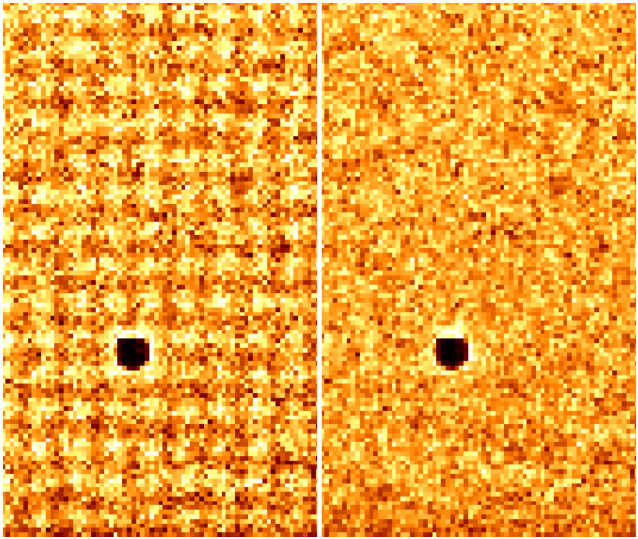}
\caption{The two images above contain the same exposure taken of a
  featureless laboratory calibration source. The image on the left was
  the one delivered by the instrument where the uniform background
  allows us to easily identify the modulo-8 pixel fixed-pattern
  structure. The image on the right is the same data after the
  fixed-pattern has been corrected. The dark area of the image is a
  dead CCD pixel.}
\label{fig:Mod8badpix}
\end{figure}

As described in Section \ref{subsec:om} the onboard event centroiding 
leads to a modulo-8 fixed pattern structure on the raw images, 
with a peak to peak amplitude of typically 10--20 per cent (see Fig.~\ref{fig:Mod8badpix}). 
It should be noted that
because the pattern originates from the centroiding process, it is 
a distortion in the positioning of counts only, 
and is therefore flux-conserving. 
The pattern repeats on a 4 arcsec scale (the CCD pixel size), so it 
is larger than the 
point source FWHM, but smaller than the generic 5.7 arcsec radius point source
and 7.5--12.2 arcsec background apertures used to generate catalog
photometry (see Section \ref{subsection:pointphotometry}). The amplitude and 
precise structure of 
the mod-8 pattern vary slowly over the detector.

Dividing out the mod-8 pattern or Fourier-filtering to remove 
it would not conserve flux and would therefore bias the 
catalogue
photometry. The approach is therefore to measure the amplitude and
form of the structure within a cell of dimensions 16$\times$16 CCD
pixels (or 64$\times$64 arcsec$^2$), which slides over the image
in order to account for variations due to detector
position and local count rates. At each step, the mean background
value is calculated with pixels containing sources and cosmetic defects
discarded using an iterative clipping algorithm if they deviate
from the mean by $\pm 6\sigma$. The assumption is made that the
remaining fixed pattern does not vary across the cell before the pixels 
are resized individually so that the background is rendered statistically
uniform. After the sliding cell has passed over the full image, the
image is resampled so that pixel sizes are once again uniform. The
process has the effect of redistributing a small fraction of photons 
between neighbouring pixels in order to reverse the residual errors in 
the on-board centroiding and minimise the fixed-pattern, while retaining
photometric accuracy.

Image backgrounds must be bright enough to provide meaningful
statistics within the sliding cell, otherwise the fixed-pattern
correction becomes Poisson noise-dominated. In such cases the
background level is too low for the fixed-pattern distortion to have
any significant impact on the detection or measurement of
sources. Therefore if a cell contains $<$1\,000 counts, corresponding
to a sensitivity limit of 25 per cent peak-to-peak pattern structure, the
correction is not performed. This situation occurs most frequently in
short UV exposures.

\subsection{Source detection}
\label{subsection:DataProcSourceDetection}

Source detection is performed in raw detector coordinates in order to
simplify the quality flagging.  Quality issues such as readout
streaks, diffraction spikes and smoke rings (see Section \ref{subsubsec:smoke})
are most easily diagnosed
and kept track of in detector coordinates before rotating,
undistorting and translating the image to sky coordinates. The
measured detector coordinates of each source are transformed to sky
coordinates in a subsequent process.

There are a large number of astronomical software packages for
performing source detection on optical images, and most either locate
sources by searching for local maxima, or by a thresholding process
\citep[see ][]{drory03}. The former method is usually the most
efficient way to detect point sources, whilst the latter one is more
suited for extended sources.  Because of the large range in OM image
background values encountered (from $<$1 count per pixel in the UV to
$>$400 counts per pixel in the optical) and issues specific to the OM
images (e.g. scattered-light features, modulo-8 patterns around bright
sources), OM images are reduced using the SAS task {\sc omdetect},
which uses both peak-finding and thresholding algorithms in order to
detect both point and extended sources as efficiently and reliably as
possible.  Details of the testing and validation of the {\sc omdetect}
source detection algorithm are given in Appendix A. A brief outline of
how it works follows.

{\sc omdetect} first constructs a background map and a noise map 
which are used to
set the detection threshold.  The background level at a particular
pixel location is computed using the median value of the pixel values
in a box 50 arcsec on a side, centred at that position, ignoring
pixels which were above a certain height above a global median and
using a simple clipping algorithm.
Each pixel of the
noise map is computed using the standard deviation of
the pixel values used at that particular point in the background map. 

The sources on an image are detected in a number of stages. The first
stage uses a peak finding algorithm to find mainly point-like sources.
A list of all the pixels above a background threshold is obtained,
and, in order of decreasing pixel value, each one is checked to see if
it could be part of a point source.  In checking the pixels in this
order, point sources are generally detected in order of decreasing
brightness and the detection of faint spurious sources in the wings of
bright sources is largely overcome.  A validation function is used
which checks the profile shapes of sources for consistency with the
point-spread function and image binning.  If a source passes the
validation tests, the pixels associated with it are identified and
used to compute the position, 
FWHM of the major axis, FWHM of the minor axis
 and
position angle using intensity-weighted moments 
(for more details see Section \ref{sec:extended}). 
The source is then
classified as either point-like or extended from a comparison of the
FWHM of the major axis with the FWHM of the point-spread
function.
If the FWHM of the major axis is larger than the FWHM of the 
point spread function by more than 3 times the major-axis FWHM uncertainty, 
the source is classified as extended; otherwise it is classified as 
point-like.
The detection process is repeated a number of times using
different validation functions and thresholds to find as many point
sources as possible. Each validation function is fine-tuned to
identify a particular kind of point source (e.g. very bright sources
surrounded by a modulo-8 distortion, sources with one or more close
neighbours, sources near an image edge).  Few sources at this stage
are classified as extended.

The next stage is to locate extended sources, and for this purpose an
image segmentation algorithm is used.  A map of the pixels above a
given height above the background map is obtained and the number of
pixels within each cluster is checked to ensure that it lies within
specified bounds; those clusters that do not are discarded.  The
remaining pixel clusters are examined using a validation function.
The validation function carries out various checks (e.g. pixel
geometry, proximity to bright point sources, axial-ratio) and if the
pixel cluster passes the checks its position, FWHM of the major axis,
FWHM of the minor axis and position angle are computed using
intensity-weighted moments. 
As in the previous stage, the source is classified as point-like
or extended according to whether its major-axis FWHM exceeds the FWHM of 
the point spread function by more than 3$\sigma$.
The process is repeated several times to
find both large and small extended sources.  A special validation
function is used to attempt to identify scattered light regions. Some
of these features can be identified by their cirrus-like nature, which
leads to a distribution of source pixels containing many holes, and by
comparing the total number of holes to the total number of cluster
pixels a decision is made whether or not the source is a
scattered-light feature. If it is a scattered-light feature then these
pixels are subsequently ignored.

A final detection stage locates very faint point
sources which may have been missed in the point-source detection stage
for various reasons (e.g. there being too few pixels to compare the
profile with that of the PSF). This uses the extended source detection
algorithm with a special validation function for faint sources. 
The image is first slightly smoothed to help locate faint, extended sources. At
the end of the source-detection stage, if any large extended sources
have been detected the background map is refined by interpolating
across these sources. Each large extended source is
then re-examined to see if it can be split into two or more
overlapping sources.  

After the source-detection passes, all sources are tested for 
credibility by the following criteria:
the centroid of the source must be contained within the
cluster of pixels associated with it;
the significance of the source (i.e. the ratio of the source counts to the 
rms background fluctuation in the same aperture) must be $> 3$ when 
measured with a 5.7 arcsec (12 unbinned pixel) radius aperture; 
if the significance in a 5.7 arcsec radius aperture 
is $< 10$ the source must also have a significance $>3$ in 
a 2.8 arcsec (6 unbinned pixel) radius aperture. 
The next stage is to perform photometry on the sources. 
For each source, the photometry procedure which is applied depends on 
whether the source was classified as point-like or extended 
according to whether its major-axis FWHM exceeds the FWHM of 
the point spread function by more than 3$\sigma$.

\subsection{Photometry of point sources}
\label{subsection:pointphotometry}

Standard, unweighted aperture photometry is performed on each point
source. Bad pixels recorded in the quality map (see Section
\ref{subsec:BadPixels}) are ignored during the pixel summation.  The
aperture is a circle of radius 5.7 arcsec (12 unbinned pixels),
and
the background is a 7.5--12.2 arcsec circular annulus. 
The size of the source aperture is chosen to allow accurate correction 
for coincidence loss (Section \ref{subsec:om}).
The background annulus is adjusted
slightly to ensure that the background aperture is
2 full binned-image pixels from the source aperture. If source apertures of
neighbouring objects encroach upon either the background or source
regions then all contaminated pixels are ignored in the aperture
summation.

If the significance of an object is $<$ 10
the source
aperture is reduced in size to a 2.8 arcsec radius circle in order to
improve the quality of the photometry. For sources with one or more close 
neighbours the aperture is also reduced in size as necessary to avoid 
overlapping source apertures, down to a minimum of 2.8 arcsec radius. 
Reduced-aperture source counts are corrected to their 5.7 
arcsec equivalent using a
curve-of-growth extrapolation of the calibrated point-spread function
(PSF) for the appropriate bandpass, stored in the Current Calibration
File. 

Before converting source counts to a physical magnitude or flux scale
it is necessary to correct for potential photon-coincidence losses
(Section \ref{subsec:om}). The brighter a source, the greater the level
of coincidence loss. The effect has been calibrated for point sources
through observations of photometric standard fields. Coincidence loss
distorts the point spread function at high countrates, so the
calibration has been determined for apertures of a fixed size of 5.7
arcsec (12 unbinned pixel) radius, 
an aperture size which minimises the effect of the distortion on 
the measured count rate.
The coincidence loss correction is
therefore applied directly to the counts within these apertures, or
for weak sources, to the counts after they have been adjusted to these
aperture sizes.
For the UVW1, UVM2 and UVW2 filters, the zeropoints in 
the Current 
Calibration File are for a 16.5 arcsec aperture, and so for these filters 
a further curve-of-growth correction is now applied to obtain count 
rates in a 16.5 arcsec aperture. 

The XMM-OM has a time-dependent sensitivity variation owing to gradual
photocathode degradation. This is calibrated by the XMM-OM instrument
team using regular calibration observations and it is recorded in the
Current Calibration File. Correction of the source counts for the
time-dependent sensitivity variation is performed after the
coincidence-loss and aperture corrections. 
Finally, the source counts are converted
to magnitudes and flux densities using the zero points and 
conversion factors stored in the Current Calibration File.
Photometric uncertainties are calculated as Poisson errors, scaled by 
the same factors (coincidence loss, aperture correction and time-dependent 
sensitivity) as the source counts. The photometric uncertainties do not 
include systematic terms related to uncertainties in the coincidence loss 
or photometric calibration. 

\subsection{Photometry of extended sources}
\label{subsection:extendedphotometry}

The aperture for extended source photometry is source-dependent and
irregular, containing all pixels associated with the object during the
source detection operations. All clustered pixels $> 2\sigma$ above the
background are considered to be the same source. Bad pixels recorded
in the quality map (Section \ref{subsec:BadPixels}) are ignored during
the pixel summation. The background level is determined from the background 
image produced during the source detection process 
(Section \ref{subsection:DataProcSourceDetection}).

Coincidence losses are calculated for each individual
pixel using a 5.7 arcsec 
aperture centred on the pixel. Both source and background 
pixels are corrected, and the source counts are obtained by 
summing all corrected pixels within the source aperture minus the
inferred summation of corrected background within the aperture.

The time-dependent sensitivity degradation correction is identical to
that applied to point sources. The same zeropoints and flux conversion
factors are used for extended sources as for point sources. The
magnitudes and fluxes quoted in the catalogue are integrated over the
whole aperture, rather than provided per unit area.

\subsection{Astrometric correction}
\label{subsection:astrometry}

The first step in obtaining sky coordinates for sources is to correct
their raw detector coordinates for image distortions resulting from
the optics and detector \citep{mason01}. This distortion correction is
stable with time and is stored in the Current Calibration File. After
the distortion correction the plate scale is linear, and the source
positions on the detector are transformed into sky positions using the
pointing information recorded by the {\it XMM-Newton} star trackers.
The {\em XMM-Newton} attitude control system maintains closely the
relative spacecraft pointing once it has settled after a slew between
one science field and the next, and in the majority of cases
spacecraft drift and jitter are small compared to the XMM-OM
PSF. However, the absolute accuracy of the star trackers is a few
arcsec, so a fine correction to the aspect solution is performed
during pipeline processing. Positions of detected sources are compared
to the USNO-B1.0 source catalogue \citep{monet03} and a single linear
correction in RA and Dec is converged upon for each image by source
matching. No correction is made to the satellite position angle (and
hence the rotation of the image) because tests indicate that such a
correction is not required.  A final aspect refinement is obtained by
repeating the source matching after merging the source tables from the
individual sub-exposures of an {\it XMM-Newton} observation.
  After USNO-registration, the astrometry of individual objects
  detected in XMM-OM images is limited by systematics in the
  distortion map, which are currently measured to be 0.7 arcsec rms
  \citep{talavera11}. A comparison of the XMM-SUSS catalogue positions
  with positions derived from external catalogues is given in
  Section \ref{subsec:astrometryvalidation}.

\subsection{Quality flagging}
\label{subsec:flags}

Sources which are suspected to originate from an artifact of some
kind, or for which the measurements are likely to be compromised in
some way, are flagged as such during the construction of the
catalogue. Table \ref{tab:flags} lists the various quality flags which
may be set during the processing. The flags are treated as bits in a
binary number, so when they are set, bits 0,1,2,3,4,5,6,7,8,9 are 
equivalent to the integer numbers 1,2,4,8,16,32,64,128,256 and 512 
respectively. For convenience the quality flags are recorded in the 
catalogue both as integer
numbers (the sum of the bit values of the flags which are set) 
and as strings of true/false logical values. The meaning of these quality
flags and criteria by which they are set are described briefly in the
following subsections, with some more detailed criteria for flags 1--4 
provided in Appendix B. We close this section with a few tips on the use of 
the flags. 

\begin{table}
  \centering
    \caption{Source quality flags within the XMM-SUSS catalogue.}
    \label{tab:flags}
    \begin{tabular}{cl}
      \hline
      \multicolumn{1}{c}{Bit} & \multicolumn{1}{l}{Quality issue}\\
      \hline
      0 & Source lies over a bad pixel\\
      1 & Source lies on or near a bright readout strip\\
      2 & Source lies on or near a smoke ring\\
      3 & Source lies on or near a diffraction spike\\
      4 & Source is bright with coI-loss-induced mod-8 noise\\
      5 & Source lies within the central region of scattered light\\
      6 & Source lies close to another bright object\\
      7 & Source lies close to or over an image boundary \\
      8 & Point source lies over an extended source\\
      9 & Point source is too compact\\
      \hline
    \end{tabular}
\end{table}

\subsubsection{Flag 0: bad pixel}
\label{subsubsec:bad}

An  object is  flagged if the  photometric aperture  used  to sum
source  counts  includes any  bad  pixels,  as  recorded in  the XMM-OM
Current Calibration File,  blank  areas  of image  caused  by  telemetry
dropouts, or isolated bright pixels.

\subsubsection{Flag 1: readout streak}
\label{subsubsec:streak}

Readout streaks occur because there is no shutter to block incident
photons during the short but finite time the CCD takes to readout.  It
is a negligible effect in sources with brightness within the useful
dynamic range of the instrument, but readout streaks from saturated
field sources will both contaminate neighbours situated along the same
readout column and cause spurious source detections.  The criteria
used to identify readout streaks are described in Appendix B. Sources 
likely to be resulting from, or contaminated by, readout streaks are flagged.
   
\subsubsection{Flag 2: smoke ring}
\label{subsubsec:smoke}

Internal reflection of light within the detector window results in an
out of focus ghost image of each source which is displaced radially by
the curved detector window. These artifacts are called smoke rings.
Smoke rings from sources of brightness bounded within the effective
dynamic range of the instrument contain negligible numbers of photons
but they are noticeable artifacts in saturated sources. Any sources
detected within 19 arcsec from the center of a potential smoke ring
could be related to, or have their photometry compromised by, the
smoke ring and are flagged as such. It should be noted that there are
circumstances in which the source responsible for generating a smoke
ring is outside the image in which the smoke ring occurs; such cases
will not be flagged.

\subsubsection{Flag 3: diffraction spike}
\label{subsubsec:diffraction}

The secondary mirror  support  vanes  give  rise  to  diffraction  
spikes  in  the
brightest  sources which provide  undesirable image  structure around
neighbouring  sources and  generate spurious  sources. Sources which are 
likely to lie close to diffraction spikes are flagged.

\subsubsection{Flag 4: bright source surrounded by coincidence-loss-induced 
modulo-8 pattern}
\label{subsubsec:mod8}

As explained in Section \ref{subsec:om}, sources with countrates
approaching 1 count per image-frame are subject to coincidence loss,
which distorts the point spread function and gives rise to a modulo-8
pattern in the region surrounding the source. The morphologies of such
sources cannot be recovered, and hence they are flagged during the
construction of the catalogue.  
  At very high countrates
  coincidence loss leads to saturation and the photometry of sources
  cannot be recovered (see Section \ref{subsec:om}). Occasionally,
  sources approaching saturation can lose counts due to integer
  wraparound in individual pixels of the raw data. Photometry of
  sources approaching the saturation limits should always be treated
  with caution.  

\subsubsection{Flag 5: central enhanced region}
\label{subsubsec:scattered}

Scattered light from the detector chamfer leads to an annular region
of background in the centre of the field of view which is enhanced by
more than a factor of 2 with respect to the background over the rest
of the field of view. Sources within 1.25 arcmin of the instrument
boresight are flagged. Note that the flag indicates that a source is
{\it within the region} corresponding to the central enhancement,
whether or not the background level is large enough for the central
enhancement to have a significant effect.

\subsubsection{Flag 6: close to bright source}
\label{subsubsec:confusion}

The structure around bright sources could lead to spurious sources
being detected. Any source within 33 arcsec of a source flagged as
bright with coincidence-loss-induced modulo-8 structure (flag 4) is
flagged as lying close to a bright source.

\subsubsection{Flag 7: image edge}
\label{subsubsec:edge}

Photometry and astrometry will be compromised if a source is partly
outside the imaged area. Sources for which any part of the photometric
aperture lies outside the field of view (including the corners of the detector)
are flagged.

\subsubsection{Flag 8: embedded within extended source (not used)}
\label{subsubsec:extended}

Photometry of point sources is complicated if they overlap an extended
source because both source and background aperture will include some
contribution from the extended source.  If any pixel of the
photometric aperture of a point source is shared by an extended source
then the point source is flagged.  
  The flag-8 algorithm did not
  set this flag for any sources retained in the final, released
  catalogue, and hence the bit corresponding to flag 8 is, in practice, 
  merely a placeholder.

\subsubsection{Flag 9: too compact}
\label{subsubsec:toocompact}

Although the XMM-OM SAS processing checks the raw data for image bits
which have been corrupted, artifacts are occasionally missed and
sources are occasionally found in the reduced images which are too
compact to be consistent with the XMM-OM point spread function. Such
sources are flagged as they are likely to be spurious.

\subsubsection{Advice on using the flags}
\label{subsubsec:flagtips}

While we recommend that the user treat any quality-flagged source with
caution, we offer here some practical advice on the use of the flags. Flag 0 is
in the majority of (though not all) cases benign because {\sc
  omdetect} uses a relatively large area to compute the moments and
photometry. Faint sources, and especially faint, extended sources with
flags 1, 2 and/or 5 are potentially spurious, but bright sources
(significance $>10$) with these flags set are likely to be valid, and
their properties robust. The central background enhancement is usually
very weak in the UVM2 and UVW2 images, so flag 5 is usually benign in
these bandpasses even for faint sources. The user should be wary of
photometry, astrometry and morphology of sources with flag 6 or 7,
regardless of source brightness. Photometry of sources with flag 4 set
is usually good provided they are not approaching the coincidence-loss
limits described in Section \ref{subsec:om}, but morpological
information for sources with flag 4 should not be trusted. Sources with 
flag 9 are potentially spurious, but this flag is rare (only 17 sources 
in the XMM-SUSS have flag 9 set).

\subsection{Merging and concatenation of the sourcelists}
\label{subsec:merging}

A sourcelist is generated for each individual XMM-OM imaging
  exposure.  For each {\em XMM-Newton} observation, the sourcelists
  from all the XMM-OM exposures, which may include multiple exposures
  through the same filter and/or exposures through different filters,
  are merged to form a single sourcelist for the observation using the 
SAS task {\sc omsrclistcomb}.  
Objects are identified as being a single unique object if they are
displaced by less than 2.0 arcsec, or 3 times the positional
uncertainty, whichever is the larger. 
The resulting source lists have
one row per source, and are the building blocks of the catalogue,
which is formed from their concatenation. The quantities listed in
each source row (position, magnitude, etc) are average quantities from
the measurements in individual exposures within the
observation. Objects which are identified in multiple {\em XMM-Newton}
observations will have multiple entries in the XMM-SUSS. These objects
are allocated the same ``SRCNUM'' identifier in the catalogue if they 
are within 2 arcsec (point sources) or 3 arcsec
(extended sources) or if they are within 3 times the position
uncertainty, of each other. 
  This final stage of concatenating the source lists and matching
  sources between {\em XMM-Newton} observations is carried out using
  software written specifically for the XMM-SUSS, that does not form
  part of the {\em XMM-Newton} {\sc SAS}.  Almost all source matches
  are within 3 arcsec, because only 0.8 per cent of the sources have
  position uncertainties larger than one arcsec. The minimum 2 or 3
  arcsec matching radii are intended to account for systematics in the
  astrometry such as the slight undersampling of the PSF by the image
  pixels, and uncertainties in the distortion correction (see Section
  \ref{subsection:astrometry}). In extended sources, there is the
  additional issue that morphologies can be different in the different
  bands, leading to a small (but real) difference in the centroid
  position.

  Choosing a matching radius is a compromise between completeness in
  matching (which improves as the radius is increased) and minimising
  the number of spurious matches (which becomes more of a problem as
  the radius is increased). The matching radii were chosen, and the
  reliability of the matching process in the {\sc omsrclistcomb} task
  was tested, using a combination of simulations and by visual
  inspection of the matched sources in real XMM-OM sky images. In the
  simulations, a master source-list file was created with a given
  number of sources at a variety of significance levels. Using this
  master source list a number (up to 1000) of new source lists were
  generated that had a random fraction of the sources of the master
  source list, and for which each of the sources is displaced by a
  random position error. A systematic offset was also introduced in 
  each source list generated to simulate the pointing uncertainty. {\sc
    omsrclistcomb} was then run with these source lists. The merged
  source list it produced was compared to a cut-down version of the 
master source list, containing only sources which were used once or more 
in the generation of source lists with position errors. 
This testing verified the source matching, and that the
  algorithm worked well, including in crowded fields.

At the end of the catalogue merging, each unique source is labelled with a
unique source number (``SRCNUM'') within the catalogue, so that each
detection of the same source is labelled with the same SRCNUM.

\section{Validation of the catalogue}
\label{sec:validation}

In this section we describe the tests which were carried out to check the 
quality and reliability of the final catalogue.
 
\subsection{Astrometry}
\label{subsec:astrometryvalidation}

\begin{figure}
\includegraphics[width=85mm]{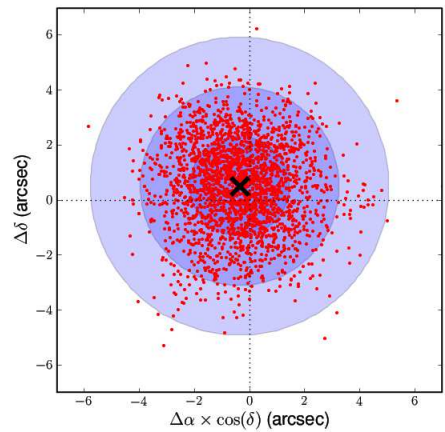}
\caption{Using source correlations between the XMM-OM source tables
  and the USNO-B1.0 catalog, systematic corrections are applied to the
  recorded XMM-OM pointings and source positions. The above plot
  summarizes the full sample of pointing offsets applied in both the
  RA ($\alpha$) and Dec ($\delta$) directions (red dots). The black
  cross is the mean correction and the blue circles represent 1-, 2-
  and 3-$\sigma$ deviations from the mean.}
\label{fig:PointingOffsetCorrection}
\end{figure}

\begin{figure}
\includegraphics[width=60mm,angle=270]{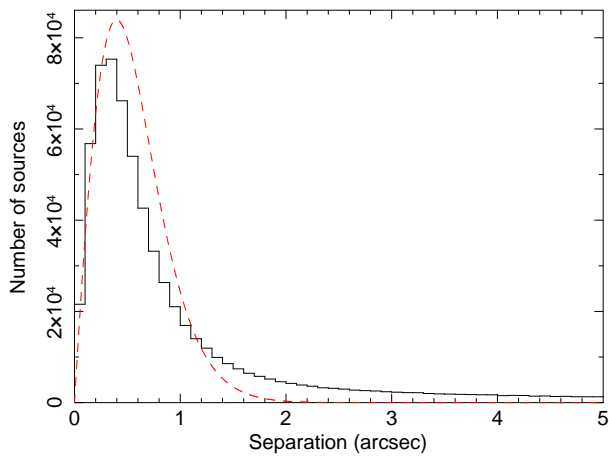}
\caption{The histogram shows the distribution of angular offsets
  between all detected sources and their closest matches in the
  USNO-B1.0 catalogue. The red dashed line is the predicted
  offset distribution obtained by summing the Rayleigh distributions
  corresponding to the distribution of position uncertainties.}
\label{fig:CorrectedAstrometry}
\end{figure}

The best-fit offsets derived from the astrometric correction process
for the XMM-OM images (Section \ref{subsection:astrometry}) are displayed
in Fig.~\ref{fig:PointingOffsetCorrection}. The
RMS dispersion of the offset distribution is 1.81 arcsec and
represents the characteristic absolute accuracy of {\it XMM-Newton}
pointing. The black cross in Fig.~\ref{fig:PointingOffsetCorrection}
represents the mean of the sample and it does not occur at the origin,
indicative of a systematic offset in spacecraft pointing,
$\langle\Delta \alpha \cos{\delta}\rangle$ = $-$0.36 arcsec and
$\langle\Delta \delta\rangle$ = $+$0.50 arcsec. 

The XMM-SUSS source catalogue includes an estimate of the 68 per cent
statistical uncertainty in the position of each source. The
uncertainties range from $0.05$ arcsec to $2.57$ arcsec, with a mean
position uncertainty of $0.68$ arcsec; they do not include any
systematic term, although systematics in the distortion correction are
known to limit the positional uncertainty to $0.7$ arcsec
\citep{talavera11}. The histogram in
Fig.~\ref{fig:CorrectedAstrometry} shows the distribution of angular
offsets between individual sources and their nearest USNO matches to a
maximum of 5 arcseconds. If the positional uncertainties in x and y
are Gaussian distributed, then we would predict the probability
distribution for the offset to be a Rayleigh distribution. The red
dashed line in Fig.~\ref{fig:CorrectedAstrometry} shows the predicted
distribution of offsets, computed as the sum of the Rayleigh
distributions corresponding to the sources and their individual
positional uncertainties. The real distribution peaks at smaller
offsets than the sum of the Rayleigh distributions, but has a tail of
associations extending beyond 2 arcseconds. 
  Investigation of sources within this tail shows that 40 per cent 
have quality flags 6 and/or 7, and so their positional
  accuracy is degraded because they are close to an image edge, or are
  affected by the modulo-8 distortion surrounding a bright source.
 Other likely contributors to
the tail include residual systematics related to the distortion map,
extended (and asymmetric) sources, for which the offset distribution
is inherently non-Rayleigh. 
The XMM-SUSS sample reaches deeper
magnitudes than USNO, so there will also be some contribution to this
tail, particularly at the largest offsets, from XMM-SUSS sources 
which are not detected in USNO and so are spuriously matched to 
unrelated USNO sources. 
Overall,
68 per cent of the offsets are within $0.8$ arcsec, and 90 percent of
the offsets are within 2 arcsec.

As a further check on the XMM-SUSS astrometry, we have cross-checked
it against the Guide Star Catalogue version 2.3 \citep[GSC2.3;
][]{lasker08}, which is independent in the sense that it was not used
in the astrometric correction process.  We have cross-matched all
sources in the XMM-SUSS with the GSC2.3, taking the closest
counterpart within a matching radius of 2 arcsec. Note that this 
cross match compares GSC positions to XMM-SUSS positions {\it after} 
the XMM-SUSS astrometry has been corrected using USNO.  We found GSC2.3
matches to 314,452 XMM-SUSS sources.  We find an average right ascension
residual $\langle \Delta \alpha \rangle = 0.06$ arcsec, with a scatter
of $0.52$ arcsec, while the average declination residual is $\langle
\Delta \delta \rangle = 0.03$ arcsec, with a scatter of $0.47$ arcsec.
The average of the residuals, in both components, is consistent with
0, indicating that there is no significant systematic offset between 
the astrometry of the XMM-SUSS and the GSC2.3. 

\begin{figure*}
\includegraphics*[width=170mm]{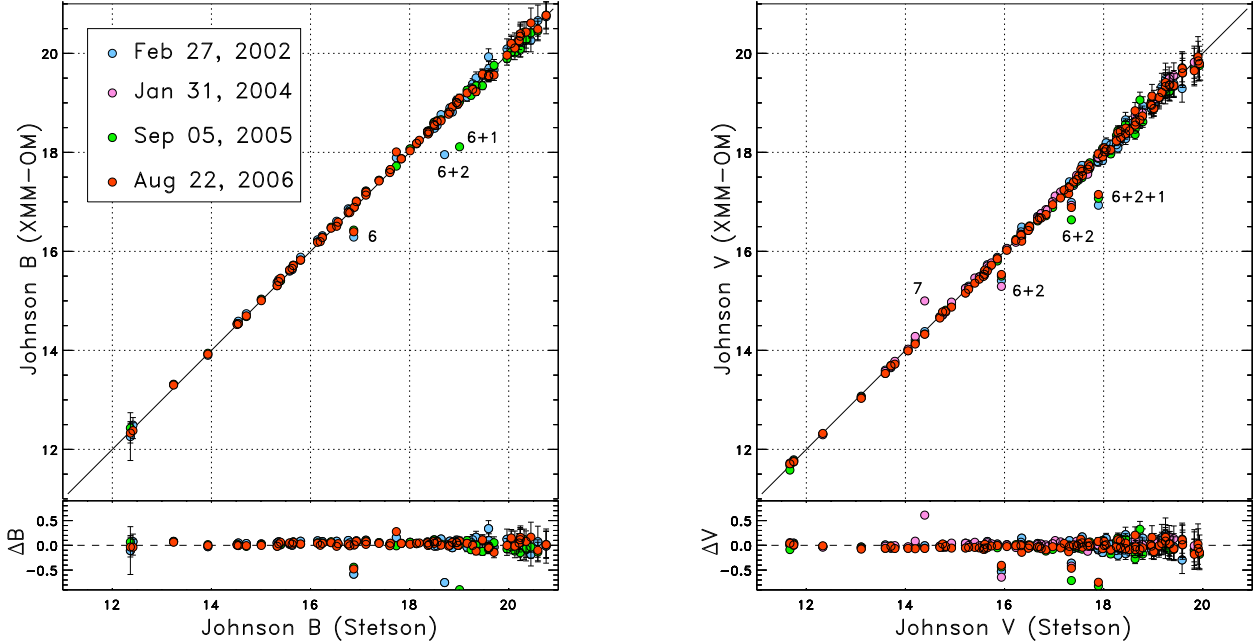}
\caption{Comparison of XMM-OM and Stetson B and V source magnitudes in
  the Landolt standard field SA95. XMM-OM magnitudes have been
  colour-corrected to the Johnson filter system. Lower panels provide
  the difference between the two measurements. XMM-OM sources were
  detected over four separate epochs. 
Photometric outliers have been labelled with the quality flags which apply 
to them (see Section \ref{subsec:flags} and Table \ref{tab:flags}).
}
\label{fig:PhotometryStandards}
\end{figure*}

\subsection{Photometric accuracy}
\label{subsubsec:PhotometricAccuracy}

Optical and UV zero-point and colour transformation calibrations
are described in \citet{antokhin01}, \citet{kirsch04} and \citet{talavera11}. 
These are based upon observations of
white dwarf spectrophotometric standards from the International
Ultraviolet Explorer (IUE; \citealt{falker87}) and Space Telescope
Imaging Spectrograph (STIS), stored in the CALSPEC database
\citep{bohlin07}. While there are no standard UV fields with which
to verify independently the accuracy of this calibration, optical
photometry can be compared against standard ground-based imaging of
photometric standard fields. Fig.~\ref{fig:PhotometryStandards}
compares XMM-OM photometry obtained with the XMM-SUSS pipeline software 
to the Stetson standards \citep{stetson00} in the SA95
field \citep{landolt92} over four mission epochs.  Two-colour
corrections \citep{talavera11} have been performed on the OM data to
adjust magnitudes to the Johnson B and V bandpasses
\citep{johnson51}. Objects with quality flags have not been screened from 
Fig.~\ref{fig:PhotometryStandards}, and are responsible for 
the obvious outliers. 

The XMM-OM and Johnson magnitudes are in good agreement; the median
offsets between the XMM-OM and \citet{stetson00} photometry for the
four epochs combined are 2 and 3 per cent in V and B bands
respectively, and the median offsets obtained from the four epochs
separately deviate from the overall medians by no more than 2 per cent. 
The stability
of the comparison over the four epochs is testament to the accuracy of
time-dependent adjustments made to the calibration in order to
compensate for detector degradation.

\begin{figure}
\includegraphics*[width=82mm]{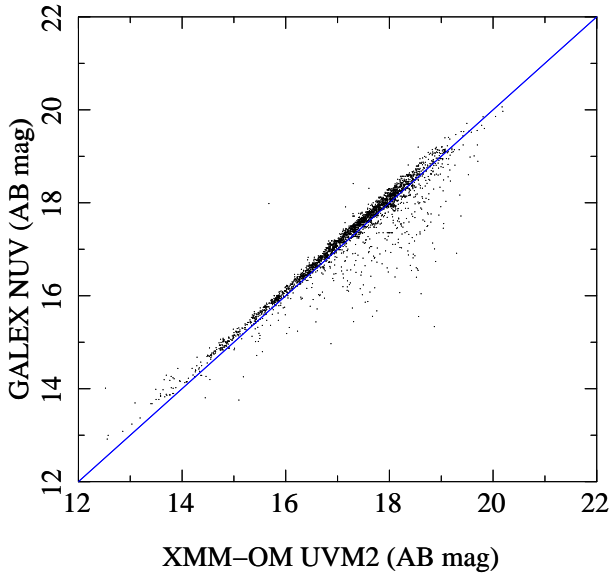}
\caption{
Comparison of XMM-OM UVM2 and {\it GALEX} NUV magnitudes. A
  good linear correlation between magnitudes in the two passbands is
  evident. The blue line corresponds to a 1:1 relation between UVM2 and NUV.
}
\label{fig:OM_GALEX}
\end{figure}

There is an absence of standard fields for photometric verification of
the UV zeropoints. As a rudimentary check on the UV photometry in the
XMM-SUSS, we compare the XMM-SUSS sample against the \galex\ Release 6
(GR6) catalogue \citep{morrissey07}.  Of the XMM-OM and {\it GALEX}
passbands, the most similar pair is XMM-OM UVM2 and {\it GALEX} NUV
(Fig.~\ref{fig:effarea}),
though we expect some
offset and scatter in photometry betweeen the two bands because the {\it GALEX}
NUV band is much broader than the XMM-OM UVM2. 
 Sources were matched between the XMM-SUSS
and {\it GALEX} samples using a cross-matching distance of 2.0
arcsec. This yields {\it GALEX} NUV magnitudes for 31\,120 XMM-SUSS
sources which are detected in UVM2 and have no quality flags. 
The comparison was restricted to objects for which the 
photometric uncertainty is less than 0.05 magnitudes in both UVM2 and 
NUV bands to limit the scatter introduced by photometric errors, 
reducing the sample to 2\,910 objects.
The
comparison between the {\it GALEX} and XMM-SUSS photometry is shown in
Fig.~\ref{fig:OM_GALEX}, and shows a strong linear relationship
between the two magnitudes over a wide magnitude range. 
We find a mean magnitude offset of
$\langle$UVM2$_{AB}-$NUV$_{AB}\rangle=-0.026\pm0.006$, though with
significant scatter in the relation (the standard deviation is 0.33),
and the distribution is asymmetric, with a long tail of objects with
large UVM2$_{AB}-$NUV$_{AB}$ colours; these are the outliers below the
1:1 line in Fig.~\ref{fig:OM_GALEX}. Inspection of these outlying
objects shows that they are in crowded regions of the sky, sometimes
in the outskirts of nearby galaxies, and are blends of multiple
sources in the GALEX NUV images. The median UVM2$_{AB}-$NUV$_{AB}$
colour will be robust to these spurious outliers, and is only 0.1.
The small offset between the median (or mean) UVM2$_{AB}$ and 
NUV$_{AB}$ magnitudes suggests good consistency between the XMM-SUSS 
and {\it GALEX} photometric systems.

\subsection{Photometric uncertainties}
\label{subsec:photometryerrorvalidation}

\begin{figure}
\includegraphics*[width=85mm]{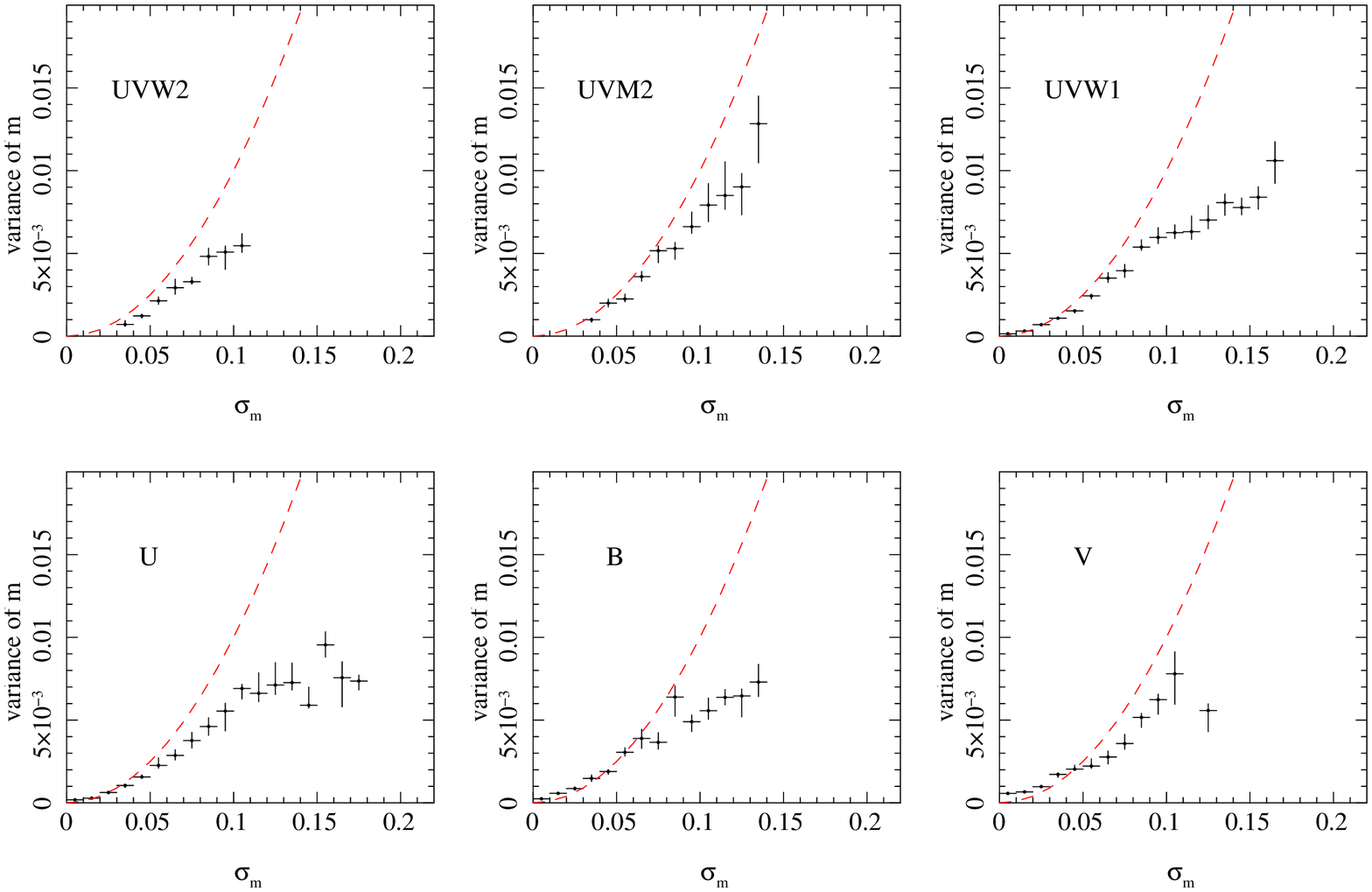}
\caption{A comparison of the median variance in photometric
  measurement, for sources observed 4 times or more, as a function of
  the mean measurement error $\sigma_{M}$, in bins of 0.01 magnitude
  in $\sigma_{M}$. Errors on the median were obtained by bootstrap
  resampling of the variance distribution in each bin. The red dashed
  line shows the expected variance given $\sigma_{M}$.}
\label{fig:PhotometryErr}
\end{figure}

A critical aspect of the catalog data is the statistical error
calculated for each photometric measurement. Many fields have been
visited more than once by {\em XMM-Newton}, and repeated measurements
of the same sources provide us with a cross-check on the statistical
uncertainties given in the catalogue. For each filter, we collate the
photometry for each source which has been detected in four or more
{\em XMM-Newton} observations with a mean significance of 10 or more,
and calculate the variance of the magnitude measurements using the
usual equation for sample variance, including Bessel's correction.
The significance cut, which is higher than that used for inclusion in
the catalogue (see Section \ref{subsection:DataProcSourceDetection}),
is included to prevent the detection limit from biasing the
variance.\footnote{If such a threshold is not applied, the photometric
  variance for low-significance sources will be artificially reduced,
  because measurements which are scattered to the fainter magnitudes
  will be below the detection threshold and so excluded from the
  catalogue.}  For each of these sources, we also calculated the mean
of the photometric uncertainties reported in the catalogue for the
relevant filter. We then determined the median of the measured
variances in 0.01 mag bins of mean photometric uncertainty. Errors on
the median were obtained by bootstrap resampling of the variance
distributions. The median variances are shown for each filter in
Fig.~\ref{fig:PhotometryErr}. The curve $y=x^{2}$ represents the
expected variance, given the photometric uncertainty reported in the
catalogue.  Apart from a small excess of variance in the V band
(equivalent to 0.02~mag additional photometric scatter), the measured and
predicted variances agree well in all filters at small photometric 
uncertainties ($\sigma_{m}<0.05$~mag).  This suggests that
systematic effects, including large scale sensitivity variations over
the face of the detector, currently calibrated to be $< 5$ per cent
\citep{talavera11}, actually contribute no more than a 2 per cent
systematic error to the photometric repeatability. For photometric
uncertainties $\sigma_{m}>0.1$~mag, the median variances lie below 
the predicted curve, and rise more
slowly with $\sigma_{m}$ than predicted. This suggests that for
$\sigma_{m}>0.1$~mag the uncertainties reported in the catalogue are
too-conservative (see Section \ref{sec:knownissues}). To summarise,
the accuracy of the photometry apears to follow the
photometric uncertainty $\sigma_{m}$ as reported in the catalogue for
$\sigma_{m}<0.1$~mag, while for $\sigma_{m}>0.1$~mag the photometric
accuracy appears to be better than reported in the catalogue.

\subsection{Extended source parameterisation}
\label{sec:extended}

The XMM-SUSS catalog provides a measure of source extent and profile
on the sky. In these diagnostics, sources are represented crudely as
two-dimensional elliptical Gaussian profiles with best-fit orientation of the
major axis relative to celestial north. Major and minor axes are
characterized by Full-Width Half-Maxima (FWHM), estimated from
Gaussian moments. To test the accuracy of the source extent
diagnostics, we compare the XMM-OM source profile properties with
those associated with optical counterparts within the Sloan Digital Sky 
Survey \citep[SDSS;][]{york00}. 

\begin{figure*}
\includegraphics[width=160mm]{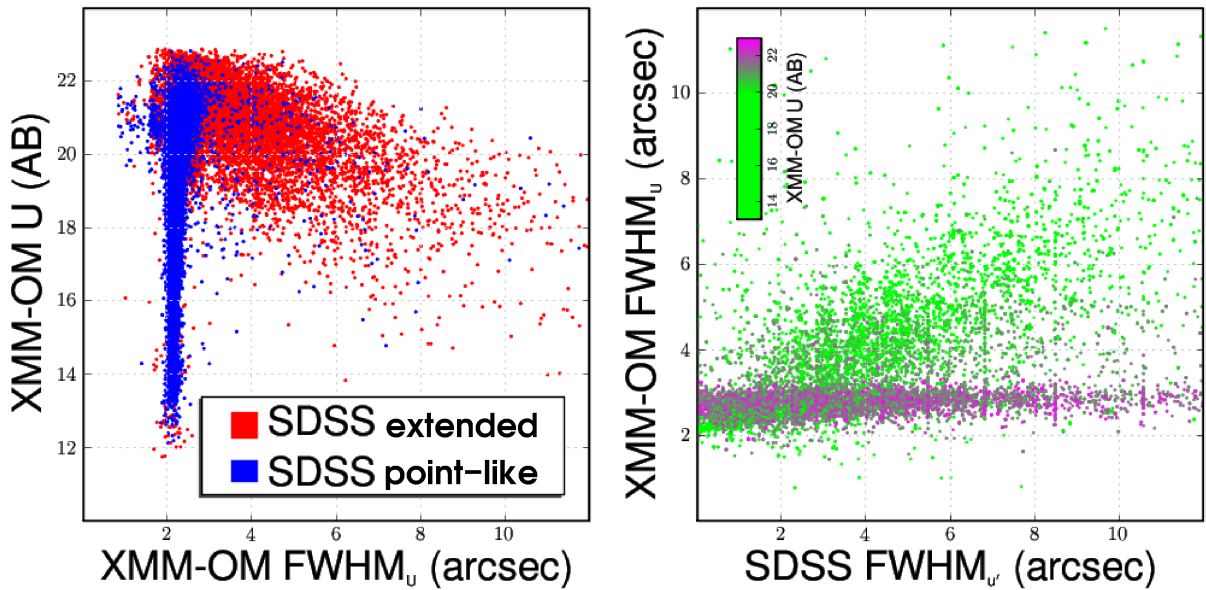}
\caption{Left panel - XMM-OM U band magnitudes of SDSS DR6
  counterparts, plotted against XMM-OM U source extent along the major
  axis. SDSS sources flagged as point-like and extended are represented by
  different colours. Right panel - comparison of XMM-OM U band source
  extent with the counterpart SDSS scale diameter from 
  fits with a de Vaucouleurs model. Colour represents XMM-OM U band magnitude.}
\label{fig:extended}
\end{figure*}

A cross-correlation of XMM-SUSS and SDSS DR6 \citep{adelmanmccarthy08} tables,
with a positional tolerance of 2 arcsec, yields 65\,579 matches. The
SDSS catalogue contains flags for source type, based upon empirical
properties. 
33\,143 sources are flagged in SDSS imaging as `STAR' (i.e. point-like), and 
30\,629 are flagged as `GALAXY', i.e. extended sources which are not 
consistent with the seeing profile of the SDSS image. 
We segregate point-like sources from extended sources and 
plot XMM-OM U band
magnitude against source extent along the major axis in
Fig.~\ref{fig:extended}. While not infallible, 
the SDSS classification is a good indication of which sources are point-like 
at the resolution of the XMM-SUSS.
Fig.~\ref{fig:extended} provides confirmation that extended sources
are being identified robustly by the SAS pipeline.  Neglecting
source-mismatches, the overwhelming majority of SDSS 
point-like sources
have
XMM-OM detections consistent with a constant FWHM of 2.2 arcsec. Note that 
the difference between this FWHM and that given in Table~\ref{tab:filters} 
is a result of the $2 \times 2$ onboard binning undersampling the image. 
There is an
anti-correlation between galaxy extent and magnitude which is a
selection effect where diffuse profile wings become increasingly
difficult to detect from fainter objects.

The right hand panel of Fig.~\ref{fig:extended} compares XMM-OM U
band 
 extent along the major axis with the corresponding SDSS
 extent, represented by twice the recorded de Vaucouleurs-law 
scale radius in the
u$^{\prime}$ band, for sources classified as extended in the SDSS.
The XMM-OM U band and SDSS u$^{\prime}$ bands are chosen
for this demonstration because they are the filter bandpasses that
most resemble each other. There are two families within this
population. There is direct correlation between bright XMM-OM sources
and their SDSS counterparts. However compared to the superior depth
and angular resolution of the Sloan survey, the XMM-OM is less likely
to detect the extended wings of weak (U $>$ 21) sources, resulting in
the second population of galaxies, recorded as low-significance, point
sources within the XMM-SUSS.

\subsection{Visual screening for quality control}

To validate the performance of the quality screening, we visually
examined 654 exposures which were chosen at random from the
observations used for the XMM-SUSS.  Sources were noted which have
quality issues but had evaded the appropriate quality flag, or which
were erroneously flagged.  Table \ref{tab:qualitystats} gives the
fractions of sources flagged with the different quality flags,
together with the statistics for missed and erroneous quality flags
derived from the visual examination. The statistics provided in Table
\ref{tab:qualitystats} should be regarded as indicative rather than
definitive: only a small fraction of the XMM-SUSS sources have been
examined, the visual screening is somewhat subjective, and the
importance and incidence of the various quality issues depend quite
strongly on filter. Nonetheless, with this caveat it is encouraging to
note that in the screened images the maximum fraction of detections,
for any quality flag, which was missed by quality flagging, or was
erroneously flagged, is 0.2 per cent.

\begin{table}
\caption{Overall fractions of detections assigned the different quality flags 
(Section \ref{subsec:flags}). The column labelled `Detections missed in flagging' 
gives the fraction of source detections identified in the visual screening which 
warranted (but lacked) a given flag. The column labelled `Detections erroneously 
flagged' gives the fraction of sources which were assigned quality flags which 
the visual screening indicated were not appropriate. 
No statistics are given for flag 8 because no sources have been set with this flag in the catalogue.
}
\label{tab:qualitystats}
\begin{tabular}{lccc}
\hline
Flag & Detections & Detections  & Detections\\
&flagged&missed in flagging&erroneously flagged\\
&(per cent) &(per cent)&(per cent)\\
\hline
0&1.1&0.001&0.0\\
1&7.0&0.2&0.06\\
2&2.6&0.2&0.04\\
3&0.2&0.2&0.01\\
4&1.8&0.2&0.06\\
5&2.3&0.003&0.0\\
6&1.5&0.2&0.005\\
7&8.2&0.04&0.0\\
8&-&-&-\\
9&0.001&0.0&0.0\\
\hline
\end{tabular}
\end{table}

\subsection{Matching of sources between wavebands and observations}

  Source confusion, where two (or more) sources occur close enough
  together on the sky that they are either blended, or mistakenly
  identified as the same source, can be an important issue in sky
  surveys. We estimated the fraction of sources for which the source
  matching may be complicated or compromised by the close proximity of
  another source, by searching the XMM-SUSS catalogue for counterparts
  at sky positions offset in declination by 30 arcsec from the real
  XMM-SUSS catalogue positions, using the matching radii used to
  cross-match sources selected in different {\em XMM-Newton}
  observations (Section \ref{subsec:merging}). This is similar to the 
  method used by \citet{bianchi11} to estimate the level of spurious matches 
  between GALEX and SDSS sources. Overall, we find that a
  counterpart is matched in 2.56 per cent of the random sky
  positions. We note that a significant fraction of these matches
  arise in the Galactic plane and Magellanic Clouds fields where the
  sky density is high. Restricting the estimate to Northern hemisphere
  fields with $| b | > 20$ deg, the fraction of random sky positions
  which are matched to a catalogue source drops to 1.0 per cent.  From
  Section \ref{subsec:astrometryvalidation} we see that the majority of
  offsets between the positions of the same sources in different
  XMM-OM images will be less than 1 arcsec, while matches to unrelated
  sources will be distributed uniformly over the source matching area,
  which is on average 16.5 arcsec$^{2}$ in area, so that $>$80 per
  cent of the mis-matches will be at distances larger than 1
  arcsec. Thus the actual number of mis-matches will be somewhat
  smaller than the frequency of matches to random sky positions,
  because the matching algorithm used in the XMM-SUSS takes the
  most-likely counterpart, which will in the majority of the cases be the
  correct counterpart. Therefore at high Galactic latitudes and
  outside the Magellanic Clouds, the likelihood of two different
  XMM-OM sources being incorrectly associated by the source matching
  algorithm is less than 0.5 per cent.

  In practice, because the point spread function of the XMM-OM is
  around 2 arcsec FWHM in the UV, source blending in the images is
  more of a problem than mis-matching at the merging stage. Indeed,
  only 0.05 per cent of the sources have a companion source in the
  catalogue within 2 arcsec. With so few pairs at small separations
  there is little scope (or need) to clean the catalogue further by
  modifying the source-matching algorithm.  Where more than one source
  contributes significantly to the flux in any one band, the source is
  likely to be flagged as extended in the XMM-SUSS. Selecting only
  point-sources is therefore likely to exclude the majority of blended
  sources.

\section{Catalogue characteristics}
\label{sec:content}

In this section we examine the characteristics of the XMM-SUSS that 
pertain to its scientific use and provide 
some general properties of the stellar and extragalactic samples.

\subsection{Source numbers and sky coverage}
\label{subsec:SkyCoverage}

There are 753\,578 sources within the XMM-SUSS counting multiple
detections within an {\it XMM-Newton} pointing as a single
source. 
Taking into account detections of the 
same source during
different pointings, the number of unique sources within the catalogue
is 624\,049. Each unique source within the catalogue is assigned a unique 
identifier (column ``SRCNUM''), with which each entry of that 
source in the catalogue is labelled. Depending on filter, 6--9 per cent 
of the sources are identified as extended, with the U and B filters having 
the largest fractions, perhaps because these two passbands have the highest 
effective areas, and are therefore the most sensitive to faint, extended 
emission. 

\begin{figure*}
\includegraphics[width=180mm]{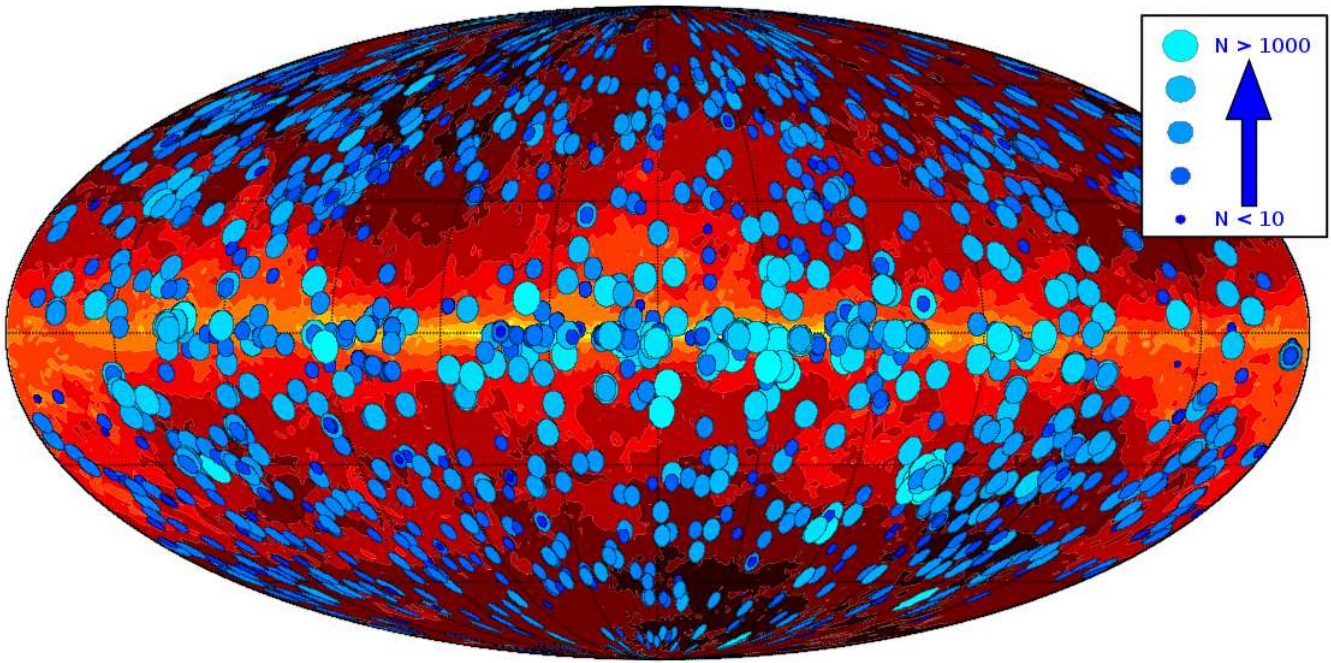}
\caption{The pointing and source number content of the XMM-SUSS
  catalogue (blue circles) mapped onto a Mollweide equal area
  projection of the sky in Galactic longitude and latitude. Size and
  colour of the circles trace the total number of UV sources $N$ in
  each pointing, after sources in different filters have been matched.
  The colour image in the background is the Infra-Red Astronomy
  Satellite (IRAS; \citealt{clegg80}) 100$\mu$m all-sky map
  \citep{schlegel98}, showing the dust associated with our Galaxy.}
\label{fig:ObservationMap}
\end{figure*}

The 2\,417 XMM-SUSS pointings are shown in Galactic coordinates 
in Fig.~\ref{fig:ObservationMap}. The size of the blue pointing symbols
increase in radius and vary in hue as the number of UV sources per 
pointing increases. 
Mirroring the wide variety of scientific experiments performed by
{\it XMM-Newton} for its Guest Observers, there are many observations at
high Galactic latitude together with clusters of pointings along the
Galactic plane, and towards the Magellanic Clouds.  
Galactic plane and
Magellanic Cloud pointings have the largest source densities.

\begin{table*}
  \centering
  \caption{Sky coverage and source statistics for each of the six XMM-OM 
    bandpasses included within the catalog. $\Omega$ is the 
    solid angle accumulated by each bandpass, $N_{\mbox{\small{tot}}}$ 
    is the number of sources found with a detection significance 
    $> 3\sigma$ and $f_{\mbox{\small{obs}}}$ is  the fraction of 
    XMM-SUSS pointings that include each bandpass.
    $N_{\mbox{\small{src}}}$ is the number of unique sources 
    detected, after repeat detections are accounted for.
    $N_{\mbox{\small{gal}}}$ is the number of sources within 
    $\pm5^\circ$ of the Galactic plane, $N_{\mbox{\small{mag}}}$ 
    is the number of sources within a cone of radius $4^\circ$, centered 
    around the Magellanic Clouds at $l = 278^\circ$, $b = -33^\circ$. 
    $N_{\mbox{\small{hi-$b$}}}$ is the number of objects at high 
    Galactic latitude, $|b| > 30^\circ$, excluding the Magellanic 
    Cloud cone.}
    \begin{tabular}{lccrrrrr}
      \hline
      \multicolumn{1}{l}{Filter} & 
      \multicolumn{1}{c}{$\Omega$ (deg$^2$)} &  
      \multicolumn{1}{c}{$f_{\mbox{\footnotesize{obs}}}$ (per cent)} &
      \multicolumn{1}{c}{$N_{\mbox{\footnotesize{tot}}}$} &
      \multicolumn{1}{c}{$N_{\mbox{\footnotesize{src}}}$} &
      \multicolumn{1}{c}{$N_{\mbox{\footnotesize{gal}}}$} &
      \multicolumn{1}{c}{$N_{\mbox{\footnotesize{mag}}}$} &
      \multicolumn{1}{r}{$N_{\mbox{\footnotesize{hi-$b$}}}$}
\\
      \hline
      UVW2 & 43.7 & 15.9 & 119\,805 &  96\,814 &  25\,097 & 50\,158 &  24\,439\\
      UVM2 & 29.0 & 19.3 & 145\,210 & 120\,510 &  25\,625 & 53\,208 &  35\,311\\
      UVW1 & 54.0 & 82.0 & 618\,266 & 521\,507 & 124\,188 & 71\,948 & 199\,934\\
      U    & 21.6 & 23.6 & 177\,569 & 146\,329 &  34\,080 &  4\,887 &  74\,815\\
      B    & 14.3 & 10.8 &  81\,191 &  62\,966 &   8\,738 &       0 &  42\,644\\
      V    & 12.6 & 10.4 &  78\,160 &  57\,498 &  27\,366 &       0 &  23\,073\\
      \hline
    \end{tabular}
    \label{tab:SkyStatistics}
\end{table*}

Table~\ref{tab:SkyStatistics} lists the total area of sky observed
through each filter (taking into account overlapping observations), the fraction of observations which include each
filter, and the number of source detections in each band. UVW1, which
has the largest throughput of the UV bandpasses, is the default filter
for science programs which contain no optical or UV science goals beyond the
serendipitous detection of field sources. Consequently, the largest
sky area is covered by UVW1, and the largest number of UV source detections
(618\,266) are made in this band. 
Indeed, many sources are detected only in UVW1.

The numbers of U, B and V sources are limited in
Table~\ref{tab:SkyStatistics} by the requirement that only those
detected sources with UV counterparts are included within the catalogue. The
source content in these filters are also significantly affected by the
rejection of fields with $> 10\,000$ source detections in order to
avoid crowding and confusion during fine-aspect correction and
aperture photometry (Section \ref{subsec:selection}). The effective
area of XMM-OM peaks in the optical and so UBV data contain a larger
fraction of crowded fields compared to the UV sample.  As seen in
Table~\ref{tab:SkyStatistics}, B and V observations of the Magellanic
Clouds in particular do not make this cut and do not appear within the
catalog. One avenue for improving future data releases of the XMM-SUSS
will be to improve the fidelity of automated crowded-field data
reduction. The fine UV imaging resolution of the XMM-OM is a great
asset which will not be fully-exploited in the XMM-SUSS until all UBV
counterparts can be incorporated within the catalog accurately.

\subsection{Magnitude and signal to noise distributions}
\label{subsec:SourceDetection}

\begin{figure}
\includegraphics[width=86mm]{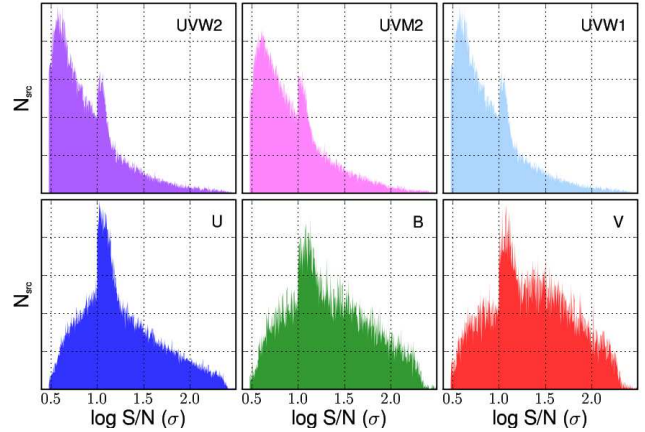}
\caption{The normalized detection significance
  distributions of sources with no quality flags contained within the
  XMM-SUSS catalog, per filter. $N_{\mbox{\small src}}$ is the number
  of sources.}
\label{fig:SignificanceDistribution}
\end{figure}

Fig.~\ref{fig:SignificanceDistribution} shows the
detection significance distribution of all the good-quality sources
contained within the catalog, by filter. 
The step-change in source numbers at $10\sigma$ is a consequence of
the requirement that sources below this significance level have to
pass an additional significance test with a smaller aperture for
inclusion in the catalogue (Section
\ref{subsection:DataProcSourceDetection}). 
 The different shapes of
the UV and optical distributions in
Fig.~\ref{fig:SignificanceDistribution} is a result of the requirement
that sources must be detected in at least one UV band for inclusion in
the catalogue.

\begin{figure}
\includegraphics*[width=85mm]{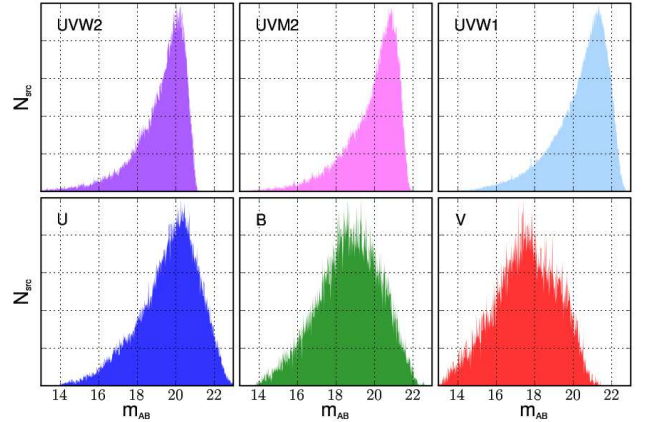}
\caption{The normalized magnitude distributions of XMM-SUSS sources. Sources with quality flags are not included in the distributions.
$m_{\mbox{\small{AB}}}$ is the AB magnitude of sources
  and $N_{\mbox{\small{src}}}$ the number of sources contained
  in each $\Delta m_{\mbox{\small{AB}}}$ = 0.01 histogram bin.}
\label{fig:MagnitudeDistribution}
\end{figure}

Fig.~\ref{fig:MagnitudeDistribution} shows the filter-dependent
magnitude distributions of sources within the XMM-SUSS. Only those
sources with no quality flags are included and the distributions are
normalized for better comparison. 
Unlike the sharp, artificial cut-offs displayed by the UV source
significance distributions at the low-brightness end
(Fig.~\ref{fig:SignificanceDistribution}), the UV faint-magnitude
distributions cut off more gradually and this is the consequence both
of non-uniform exposure lengths and time-variable background levels,
resulting in a range of image depth. 
  Magnitude completeness limits thus vary from field to field, but
  conservative limits at which all fields are essentially complete can
  be taken as the magnitudes that corresponding to a signal to noise
  ratio of 10 in an 800s exposure. Using this criterion, we estimate
  that the survey is complete in all fields down to AB magnitudes of
  \mab\ = 18.1, 18.8 and 19.4 in UVW2, UVM2 and UVW1 respectively.
  The AB magnitude distributions peak at much fainter magnitudes than
  these conservative limits: \mab\ = 20.2, 20.9 and 21.2 in UVW2, UVM2
  and UVW1 respectively.  For comparison, the \galex\ All-sky Imaging
  Survey (AIS) reaches \mab\ = 19.9 and \mab\ = 20.8 in the FUV and
  NUV filters respectively, while the \galex\ Medium Imaging Survey
  (MIS) reaches FUV and NUV depths of \mab\ = 22.6 and 22.7,
  respectively \citep{morrissey07}.

\begin{figure}
\includegraphics*[width=85mm]{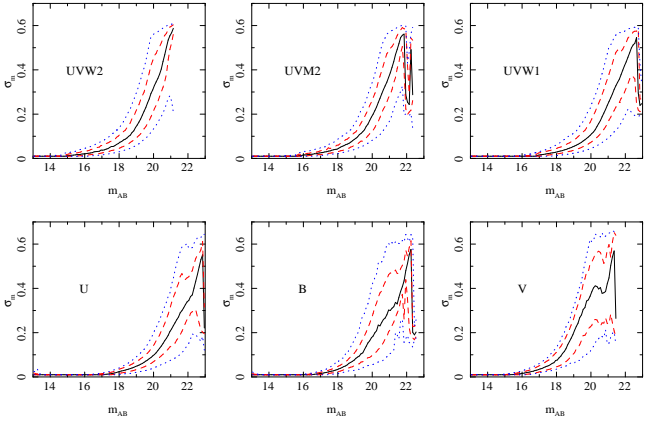}
\caption{The distribution of magnitude uncertainties in the XMM-SUSS 
as a function of passband and magnitude. The solid black line shows the 
median error, the dashed red lines enclose 68 per cent of the distribution, 
and the blue dotted lines 95 per cent. The distributions were calculated in 
bins of 0.1 magnitude.}
\label{fig:PhotometryRMS}
\end{figure}

In Fig.~\ref{fig:PhotometryRMS} we show the distribution of magnitude
uncertainties as a function of magnitude.  The solid line in each plot
gives the median uncertainty, while the dashed and dotted lines
indicate the 68 and 95 per cent limits of the distribution at
each magnitude. 
To a median uncertainty of 0.1 mag, the UVW1 and U bands probe the 
faintest AB magnitudes (19.9 and 20.0 mag respectively).
The photometric uncertainties increase significantly
to fainter magnitudes, although at the faintest magnitudes the reported
uncertainties may be over-conservative (see Section
\ref{subsec:photometryerrorvalidation}). 
  At the very faintest magnitudes the shallowest (800--1000s)
  exposures no longer contribute any sources and the catalogue becomes
  dominated by sources which are detected in the longest exposures. As
  a result, a decrease in the median photometric uncertainty is seen
  in most bands at very faint magnitudes.

\subsection{Time sampling}
\label{subsec:TimeSampling}

\begin{figure}
\includegraphics*[width=86mm]{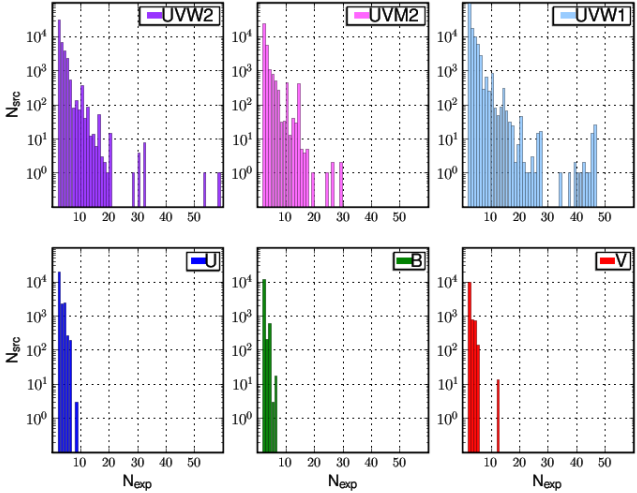}
\caption{Distribution of source re-detections during single
  pointings. $N_{\mbox{\small{exp}}}$ is the number of images
  taken during a single pointing in which a particular source has been
  detected.$N_{\mbox{\small{src}}}$ is the number of sources
  detected $N_{\mbox{\small{exp}}}$ times. Objects which are
  detected only once, $N_{\mbox{\small{exp}}}$ = 1, and sources
  with quality flags are excluded from the sample.}
\label{fig:RepeatExposures}
\end{figure}

XMM-SUSS source properties can be sampled on three distinctive time
scales. Multiple exposures through a particular filter are often
obtained during one pointing. Depending on the length of a spacecraft
visit, they provide sampling on timescales typically between hours and
a day. The longest observations are curtailed by radiation constraints
at the perigee of the 48 hour spacecraft orbit and individual
exposures are limited typically to durations $<$ 5 ksec in order to
avoid potential memory corruption after cosmic ray hits. Segregated by
filter, Fig.~\ref{fig:RepeatExposures} illustrates the number
distribution of repeat detections within individual pointings. The
differences between UV and optical distributions are a consequence of
data selection, where only those optical sources with UV counterparts
within the catalogue are selected for inclusion. Generally speaking,
if a large number of U, B or V images are being obtained, then an
observing strategy has been tailored by the investigator in order to
monitor variability in the optical bands, and no UV images will have
been taken. For sources with multiple detections in the same filter
within an {\em XMM-Newton} observation, XMM-SUSS contains the number
of detections, the $\chi^{2}/\nu$ for a constant countrate fitted to
the individual measurements 
($\nu$ is the number of degrees 
of freedom in the fit),
and the maximum deviation from the median
countrate in units of sigma. These latter two measurements allow the
selection and identification of time-variable sources in the
catalogue.
  For a non-varying source, $\chi^{2}/\nu$ is expected to be
  approximately 1, while sources with significant variability will be
  inconsistent with the constant model, and so will have larger
  $\chi^{2}/\nu$.

\begin{figure}
\includegraphics[width=85mm]{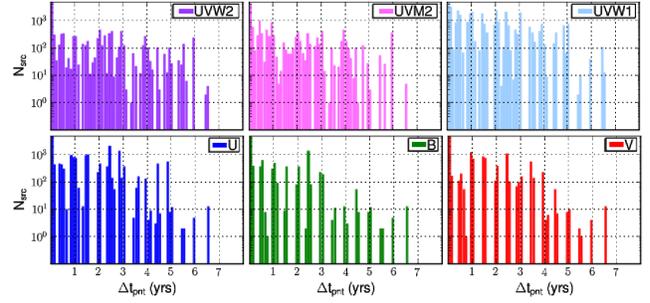}
\caption{The number of sources $N_{\mbox{\small{src}}}$ which are
  observed in multiple spacecraft pointings against the interval
  between the first and last detections of the source, $\Delta
  t_{\mbox{\small{pnt}}}$, in units of years. 
  Only sources without quality flags are included.}
\label{fig:RepeatPointingsGaps}
\end{figure}

The second timescale for source sampling is of the order of days, 
and is due to observations which are repeated within a few {\it
  XMM-Newton} orbits, typically to build up exposure times which are
longer than can be accomodated in a single orbit. The third timescale
for source sampling is of the order of years, and is associated with
re-pointings towards old fields, either serendipitously or as part of
a monitoring program. Fig.~\ref{fig:RepeatPointingsGaps} presents the
number of sources detected during more than one pointed observation,
and the number of pointings in which each source is detected, against
the time interval covered by the pointings.  The time domain for
multiple detections included in the catalog spans slightly over 8
years.  The biannual sampling apparent in
Fig.~\ref{fig:RepeatPointingsGaps} is the result of seasonal
spacecraft pointing constraints, imposed by the need to maintain the
sun angle normal to {\it XMM-Newton's} solar panels.

The three variability timescales provide the means to search for a
variety of source populations. For example, the short timescale is
sensitive to accreting binaries and coronal flares in young stars. The
long timescale is suitable for supernova searches and active galaxy
detection. Table~\ref{tab:revisits} summarizes the total number of
sources with temporal information provided within the XMM-SUSS
catalog.

\begin{table}
  \centering
  \caption{Counting statistics for sources detected in multiple 
    images. $N_{\mbox{\small{exp}}}$ is the total number of sources 
    detected in more than one exposure through a specific filter 
    and during a single spacecraft pointing. 
    $f_{\mbox{\small{exp}}}$ is the fraction of sources
    detected more than once per pointing per filter. 
    $N_{\mbox{\small{obs}}}$ is the total number of sources 
    detected multiple times during different spacecraft pointings. 
    $f_{\mbox{\small{obs}}}$ is the fraction of unique 
    sources detected during different pointings.}
    \begin{tabular}{@{}lrcrc@{}}
      \hline
      \multicolumn{1}{l}{Filter} & 
      \multicolumn{1}{c}{$N_{\mbox{\footnotesize{exp}}}$} & 
      \multicolumn{1}{c}{$f_{\mbox{\footnotesize{exp}}}$ (per cent)} & 
      \multicolumn{1}{c}{$N_{\mbox{\footnotesize{obs}}}$} & 
      \multicolumn{1}{c}{$f_{\mbox{\footnotesize{obs}}}$ (per cent)}\\
      \hline
      UVW2 &  46,031 & 38.4 & 11,657 & 12.0\\
      UVM2 &  33,660 & 23.2 & 13,010 & 11.8\\
      UVW1 & 137,309 & 22.2 & 58,280 & 11.2\\
      U    &  24,627 & 13.9 & 18,625 & 12.7\\
      B    &  12,732 & 15.7 &  9,702 & 15.4\\
      V    &   1,724 & 15.0 & 12,022 & 20.9\\
      \hline
    \end{tabular}
    \label{tab:revisits}
\end{table}

\subsection{Source colours}
\label{subsec:colours}

\begin{figure*}
\includegraphics[width=140mm]{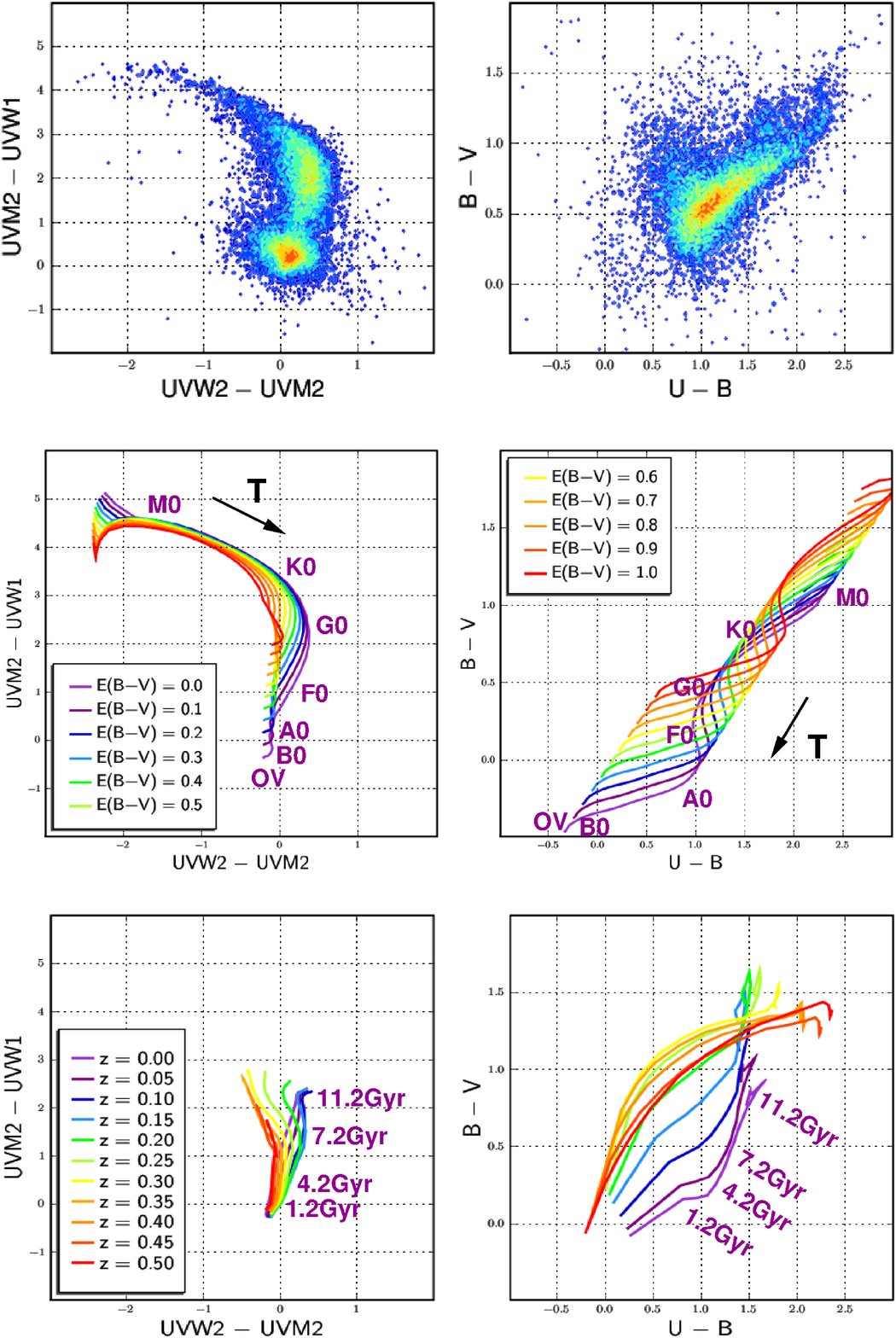}
\caption{Top panel: UV and optical colour-colour diagrams of XMM-SUSS
  sources. The density scales are logarithmic. Sources with quality
  flags have not been included in the distributions. Middle panel: UV
  and optical colours derived from ATLAS9 $\log{\left(g\right)}$ =
  5.0, solar abundance stellar spectra models behind a family of
  Galactic dust columns where $E\left(\mbox{B$-$V}\right)$ = 0.0--1.0
  and $R_{\mbox{\small v}}$ = 3.08. The arrows indicate the direction
  of increasing stellar temperature along the tracks, and 
  spectral types are indicated on the zero-reddening tracks. Bottom panel:
  galaxy colours from the E/S0 evolutionary models of
  \citet{roccavolmerange88} with a range of ages from 0.2 to 13.2 Gyr 
  at redshifts of 0.0 $< z <$ 0.5. Ages are indicated on the 
  zero-redshift tracks.
}
\label{fig:colourcolour}
\end{figure*}

For XMM-SUSS sources detected through three or more filters we can produce
colour-colour distributions. The two examples presented in the upper
panels of Fig.~\ref{fig:colourcolour} display the UVW2$-$UVM2 vs
UVM2$-$UVW1 plane and the U$-$B vs B$-$V plane, both with distinctive
sample structure. No attempt has been made to de-redden or K-correct
the colours for dust extinction or cosmological redshift. 

\subsubsection{Stellar colours}

To examine the colours of stars in the XMM-OM bandpasses we make use of 
synthesized stellar spectra, computed from 910--10\,000\AA\ and 
binned to 20\AA, provided by the ATLAS9 project
\citep{castelli03} using \citet{grevesse98} abundances. 
In the middle panel of 
Fig.~\ref{fig:colourcolour} we show the ATLAS9 \logg\ = 5.0 stellar
sequence, extinguished with a range of 
Galactic dust columns, and folded through the XMM-OM
effective area curves (Fig.~\ref{fig:effarea}). The
minimum photospheric temperature represented is 3\,000K, the hottest
is 50\,000K. We use the analytical approximation to the
Galactic dust extinction curve described by \citet{pei92} and 
characterize the dust columns by the colour excess \ebv. 
The arrows indicate the approximate direction of increasing
photospheric temperature along the tracks. In optical colours, the
stellar tracks are easily understood with stars revealing ever-bluer
colours, right-to-left and top-to-bottom as we travel from cool M to
hot O stars. The UV colours reveal a different trend however: cool
stars are located in the top left of the plot, with UVW2-UVM2 colours
increasing (i.e. moving to the right) as temperature increases 
from M stars to hotter types
before turning a corner at approximately solar spectral types. This
cool-star trend is a consequence of them having negligible UV flux 
while the red wing of the UVW2 transmission 
curve\footnote{Longward of 3\,000\AA, both the UVW2 and UVM2 filters have 
$< 0.1$ per cent transmission, and hence their red tails are too weak to be 
seen in Fig.~\ref{fig:effarea}} extends deeper into the
optical bands than that of 
UVM2 \citep{talavera11}. 
Consequently more photons are
collected from K and M stars through UVW2 than UVM2.

UVM2$-$UVW1 colours evolve monotonically towards larger values as dust
extinction increases. UVW2$-$UVM2 colours do not and there are two
contributary factors to this. The first is that the broad 2\,175\AA\
graphite feature has most effect in the UVM2 band. Increasing
extinction will deplete UVM2 photons faster than UVW2
photons. Secondly, cool stars emit very few photons in the UV
bands. Those photons which are detected from M and K stars leak in
through the red wings of the filters reponses, the UVW2 filter being
more efficient than UVM2 in these wings. There is degeneracy in the UV
colours between M dwarfs and for example highly reddened G stars so Galactic
extinction must be taken into account before identifying cool stars, but
critically there is little degeneracy between the effects of dust
extinction and the intrinsic colours of hot stars.

The UV stellar loci shown in the middle panel of
Fig.~\ref{fig:colourcolour} are mirrored faithfully in the real
catalogue shown in the top panel. The distribution of sources
indicates that the bulk of the stellar population shown in this panel
are stars of spectral-type G or earlier, with moderate interstellar
reddening ($E({\rm B}-{\rm V}) < 0.3$). This is not surprising,
because XMM-SUSS sources must be detected in all 3 of the XMM-OM UV
bandpasses to appear on the UV colour-colour diagram.

Predictably, the effects of dust extinction on the optical colours,
where the extinction curve is featureless and the wings of the filters
are not an important issue, are more simple. As the B$-$V colour
excess increases, the U$-$B colour becomes more red. Clearly, owing to
several knees in the stellar curve, there is more significant
extinction degeneracy in optical colours than in UV colours in the
region containing the majority of optical XMM-SUSS sources.

\subsubsection{Galaxy colours}

\begin{figure*}
\includegraphics[width=170mm]{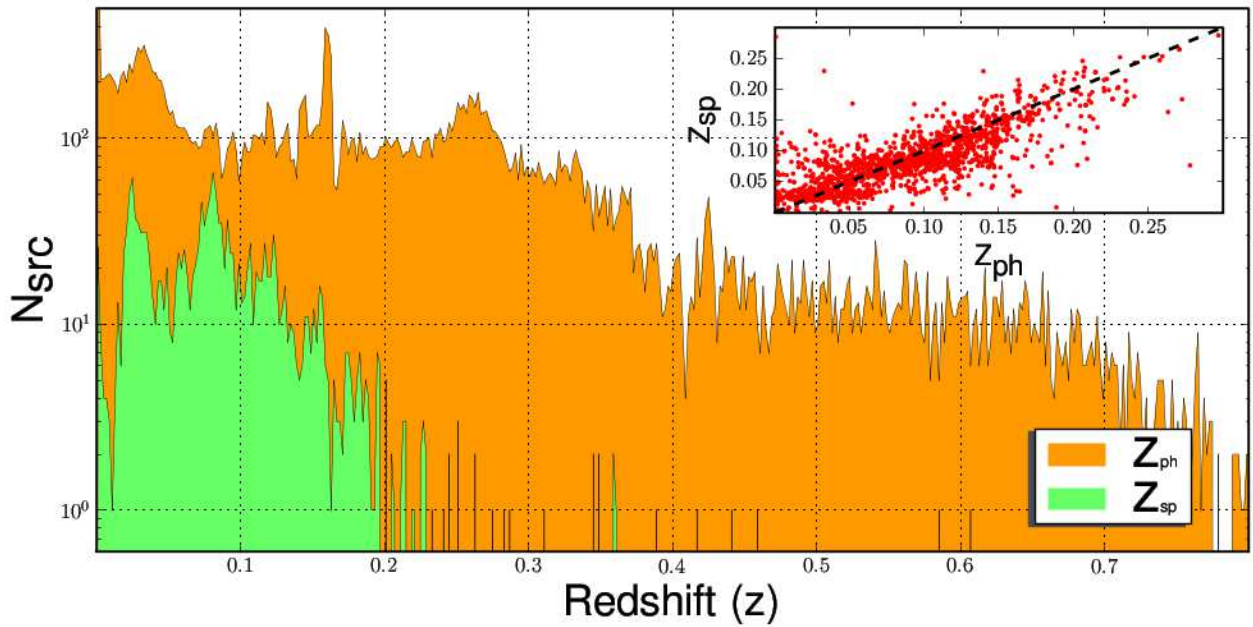}
\caption{Distribution of SDSS photometric (orange) and spectroscopic (green)
  redshifts of SDSS DR6 counterparts to XMM-SUSS sources. The histogram is sampled in bins of width $\Delta z$ = 2
\dex{-3}. The inset
  provides the photometric-to-spectroscopic distribution of sources
  where a redshift has been measured by both methods.}
\label{fig:redshift}
\end{figure*}

To examine the colours of galaxies in the XMM-OM bandpasses we have
used synthetic galaxy spectra from the template library of
\citet{roccavolmerange88}. The templates have an
exponentially-declining star formation law (the template family
``UV-cold E/S0'' in \citet{roccavolmerange88}) and a range of ages
from 0.2 to 13.2 Gyr, i.e. from strongly star-forming to old-star
dominated. The spectra were redshifted out to $z=0.5$ in discrete
steps and folded through the XMM-OM bandpasses. The synthetic tracks
so produced are shown in the bottom panels of
Fig.~\ref{fig:colourcolour}, colour coded by redshift.

In the UV colour-colour plot, the tracks corresponding to different
redshifts have a larger extent in the UVM2$-$UVW1 colour than in
UVW2$-$UVM2. They separate at the red end of the sequence, but
converge at the blue end: blue, star-forming galaxies at all redshifts
out to $z=0.5$ have UV colours UVM2$-$UVW1 $\approx$ UVW2$-$UVM2
$\approx$ 0. In the top-left panel of Fig.~\ref{fig:colourcolour} it
is seen that there is a dense cloud of XMM-SUSS sources, corresponding
to blue, star-forming galaxies, at this location in the UV
colour-colour plane.  
  The association of this cloud with galaxies is confirmed when we
  note that 85 per cent of the sources in this plot, which are flagged
  as extended in UVW1, lie in this cloud, with UVM2$-$UVW1$<1$, while
  the majority of point-like sources have UVM2$-$UVW1$>1$. However, it
  should be noted that the hot end of the stellar locus extends into
  this same region of the UV colour-colour plane as the galaxies, so
  that the UV colours alone cannot be used to separate stars and
  galaxies as reliably as the combinations of UV and SDSS colours used
  by \citep{bianchi07,bianchi11}.

The optical colours of the synthetic galaxies evolve towards the
top-left of the U$-$B vs B$-$V diagram in the interval 0.00 $< z <$
0.35 as the Balmer break moves through the B band. At redshifts $z >$
0.35 the Balmer break moves into the V band and the trend is
reversed. The spur of XMM-SUSS sources in the top-right panel of
Fig.~\ref{fig:colourcolour} with U$-$B $\approx$ 0.7 and
$0.5<$B$-$V$<1.0$ suggests that many XMM-SUSS galaxies have $z > 0.2$.

\subsection{Redshift distribution}
\label{sec:redshift}

The redshift distribution of XMM-SUSS galaxies bears heavily on their
utility for extragalactic science. The redshift distribution can be
obtained by cross-correlation of the XMM-SUSS with galaxies from the
SDSS DR6 catalogue \citep{adelmanmccarthy08}, 
which includes many spectroscopic redshifts, and a
much larger number of photometric redshifts. The catalogues were
matched by searching for the nearest clean XMM-SUSS source within 2
arcsec of an SDSS galaxy.  The redshift distributions of XMM-OM
sources with SDSS DR6 galaxies are shown in
Fig.~\ref{fig:redshift}. It suggests that the majority of XMM-SUSS
galaxies have $z<0.4$, but with a significant minority lying at
$0.4<z<0.8$.  The spectroscopic redshift limit at $z$ = 0.2 is a
systematic feature of the SDSS rather than intrinsic to the XMM-SUSS
sample. It is not clear how accurate the SDSS photometric redshifts are for
UV-selected sources at the sample limit of $z \sim$ 0.8, although the
correlation between photometric and spectroscopic SDSS redshifts
within the inset of Fig.~\ref{fig:redshift} indicates that
photometric redshifts can be trusted to within a few 10$^{-2}$ over
0.0 $< z <$ 0.25.

\subsection{Synergy with 2XMM}
\label{sec:2xmm}

As the XMM-OM data used to construct the XMM-SUSS were obtained
simultaneously with X-ray imaging data from the European Photon
Imaging Camera instruments \citep[EPIC; ][]{turner01,struder01} the
XMM-SUSS is a useful resource for studying the UV and optical
properties of X-ray sources. In this context, the 2XMM catalogue of
X-ray sources detected serendipitously in EPIC images \citep{watson09}
is the natural X-ray catalogue to pair with the XMM-SUSS and some
results from matching the two catalogues have already been 
published \citep{vagnetti10}. 

The current version of XMM-SUSS does not contain cross-references to
X-ray sources in 2XMM, so the user will have to carry out his or her
own cross matching, the parameters of which will depend on the details
of the scientific investigation being carried out. To examine the
potential sample size for joint 2XMM -- XMM-SUSS studies, we have
applied a simple cross-match between 2XMM-DR3 catalogue and XMM-SUSS
using a maximum distance of 5 arcsec. 
We obtain 14\,275 matches, and we estimate\footnote{We estimate the
  spurious match rate by cross matching the two catalogues at 5--10
  arcsec offsets, noting that almost all genuine counterparts will be
  offset by less than 5 arcsec \citep{pineau11}.}  that 90 per cent of
these will be genuine associations.

\section{Comparison with OMCat}
\label{sec:omcat}

An alternative catalogue of XMM-OM sources \citep[OMCat,][]{kuntz08},
constructed and released independently of the XMM-SUSS, is available
through the HEASARC and the Multimission Archive at STScI (MAST). Here
we outline the most important differences between XMM-SUSS and OMCat.

In scope, the two catalogues differ significantly. The XMM-SUSS is a
catalogue of UV-detected sources, i.e. it contains only sources
detected in one or more of the UVW1, UVM2 and UVW2 bandpasses, whereas
the OMCat is a catalogue of XMM-OM sources detected in any of the
XMM-OM optical and UV bandpasses. The OMCat was constructed using 
{\em XMM-Newton} observations available in the 
public archive on 1 September 2006, 
a somewhat earlier cutoff date than the XMM-SUSS bearing in mind the 
1-year proprietary period.
Obvious problem fields were screened out of the XMM-SUSS 
(Section~\ref{subsec:selection}), but were not excluded from the OMCat.
The OMCat contains a larger total
number of sources (947\,638) than the XMM-SUSS, but only around half
as many UV-detected sources (364\,741) as the XMM-SUSS. 
The increased number of UV sources in the XMM-SUSS relative to the
OMCat partly results from the expansion of the archive and partly from
improvements in the source detection chain between the construction of
OMCat and the XMM-SUSS: the peaks in the UV magnitude distributions of 
XMM-SUSS sources are more than half a magnitude fainter than those in OMCat.

Both the XMM-SUSS and OMCat were generated primarily using the {\it
  XMM-Newton} SAS tasks to process XMM-OM imaging data. The OMCat was
produced using the tasks in SAS version 6.5.0, whereas the XMM-SUSS
was produced using the tasks in SAS version 8.0. Significant
improvements were made between these two versions, in part driven by
the requirements for the XMM-SUSS catalogue. These improvements
include a more robust detection scheme for sources close to the limit
of sky background, better discrimination between point-like and
extended sources, correction for the time-dependent degradation of the
XMM-OM sensitivity, more comprehensive quality flagging and a higher
success rate (90 per cent) for refined aspect corrections. Notably,
for the construction of the OMCat a bespoke aspect correction process
was developed and employed to achieve acceptable astrometry in the
catalogue. For the XMM-SUSS, no further aspect correction was required 
beyond the refined aspect corrections derived within the SAS 
version 8.0 processing. 

\section{Known issues with the XMM-SUSS and plans for improvement}
\label{sec:knownissues}

We briefly describe a number of known problems and issues with the
XMM-SUSS, and improvements we are planning for future versions of the
catalogue. The XMM-OM instrument team intends to improve and enlarge
the XMM-SUSS on a regular basis, depending on available resources.

\subsection{Very extended sources}

Very large extended sources which occupy a significant fraction of the
XMM-OM field of view (including for example M\,31 and the Crab
nebula), were not systematically screened from the input data for the
XMM-SUSS. The XMM-OM source detection algorithm was not designed to
deal with such large sources, which it treats as non-uniform
background, and a large fraction of the detected sources in these
images are spurious. 
The resultant source lists are not, however, crowded enough to be 
excluded from the catalogue by the source density criterion described 
in Section \ref{subsec:selection}.
We intend to screen such fields from future
versions of the catalogue.

\subsection{Photometry of point and extended sources}

Sources which are marginally resolved may be identified in some
observations as extended, but be classified as point-like in other
observations.  Because photometry is computed in a different manner
for point and extended sources, this can lead to inconsistent
photometry between observations, when a source is not intrinsically
variable.  We advise users of the catalogue to check that sources
which appear to vary between observations have consistent
morphological classifications in the different observations before
concluding that they are variable. In future versions of the
catalogue, we intend to provide aperture photometry for all
sources, in addition to extended-source photometry for extended
sources, to rectify this issue.

\subsection{Lower limits for non-detections are not provided}

In the present catalogue, when a source is not detected in a
particular filter it is not possible to distinguish the cases in which
the source is not detected because it is too faint in that filter,
because no exposure was taken in that filter, or because the exposures 
in that filter were not used in the catalogue due to source crowding. 
Furthermore it
would be useful to provide the specific background limit for
sources which are not detected in some filters. However, the
functionality to perform aperture photometry over the positions of
these non-detected sources is not currently within the pipeline. This
functionality is a desirable feature for future versions of the
XMM-SUSS.

\subsection{Over-conservative photometric uncertainties at 
faint magnitudes}

As shown in Section \ref{subsec:photometryerrorvalidation} the
photometric scatter obtained from multiple observations of sources
falls below the measurement errors for measurement errors larger than
0.1 mag in all filters, implying that photometric uncertainties
$>0.1$~mag reported in the catalogue are probably larger than the real
photometric uncertainties. It follows that the variability indicators
($\chi^{2}/\nu$ amd maximum deviation in units of $\sigma$; columns
80--91, described in Appendix C) will be underestimated for objects
which have photometric errors $>0.1$~mag. We intend to diagnose and
rectify this problem before the next release of the catalogue.

\subsection{Detections of the same source in different observations are 
sometimes not matched}

We are aware of some cases in which a single source has been detected
in more than one {\em XMM-Newton} observation, but is treated as
though it were more than one source within the catalogue (i.e. the
different detections are assigned different SRCNUM values). The cause
of this problem is under investigation.

\subsection{Accuracy of coincidence loss in extended sources is unquantified}

Corrections for coincidence loss are essential for photometry of
bright sources.  Using photometric standards the effect has been
well-calibrated in point sources, but the calibration has not been
tested upon bright extended sources. This is rarely an issue for
extended sources in the UV, because the coincidence loss correction is
usually small, but for the optical pass bands it can be
significant. Events from an extended source will have a different
positional distribution compared to a point source, and hence the
coincidence loss will differ. The coincidence-loss correction of
extended sources therefore comes with an unquantified systematic
error.

\subsection{Area over which photometry is integrated for extended 
sources is not given}

Information on the area over which the flux is integrated in
extended-source photometry is not preserved in the catalogue. Such
information may be useful when comparing the XMM-OM photometry to
photometry from other sources. As such, it would be advantageous to
provide this information in a future version of the catalogue.

\subsection{Other planned improvements to future releases}

The current version of the XMM-SUSS was obtained by source-searching
individual XMM-OM exposures and merging the resulting source
lists. Functionality to aspect-correct and stack the XMM-OM images
prior to source detection has now been built into the XMM-OM SAS
tasks, and will be used in future versions of the
catalogue. Furthermore, a significant number of {\em XMM-Newton}
observations have been performed since the cut-off date for the
current XMM-SUSS catalogue. Substantial increases in both sky area and
depth are therefore anticipated in the next version of the catalogue.

\section{Conclusions}
\label{sec:conclusions}

We have described the construction, validation, and characteristics of
the XMM-SUSS, a catalogue of UV sources detected with the XMM-OM.  The
catalogue contains source positions and magnitudes in up to 6 UV and
optical bands, profile diagnostics and variability statistics. The
XMM-SUSS contains 753\,578 UV source detections which, taking into
account repeat observations of the same patches of sky, relate to
624\,049 unique objects. Taking into account overlaps between {\it
  XMM-Newton} observations, the sky area covered is 29-54 deg$^{2}$
depending on UV filter. The catalogue includes observations at a wide 
range of Galactic latitudes, including the Galactic plane. 
In terms of depth, the catalogue is typically deeper than the {\it GALEX} 
All-sky Imaging Survey: the AB magnitude distributions 
peak at \mab\ = 20.2, 20.9 and 21.2 in UVW2, UVM2 and UVW1 respectively. 
  The magnitude limits are not uniform over the survey, depending on
  background level and exposure time. However, we estimate that all
  fields are complete for magnitudes brighter than \mab\ = 18.1, 18.8
  and 19.4 in UVW2, UVM2 and UVW1 respectively.
We show that the catalogue is rich in early-type 
stars, and star-forming galaxies out to a redshift of 0.8.
The catalogue has been extensively tested for quality in 
astrometry, photometry and reliability. The XMM-SUSS has significant 
potential for
science based on temporal source variability on timescales of hours to
years, because a large fraction of sources (38 per cent in UVW2, 23
per cent in UVM2 and 22 per cent in UVW1) have been observed multiple
times through the same filter within an {\it XMM-Newton} pointing, and
a significant fraction of sources (12 per cent in UVW2 and 11 per cent
in UVM2 and UVW1) have been detected in multiple {\it XMM-Newton}
pointings. 
The XMM-SUSS catalogue provides a useful resource for a wide range of
scientific applications, whether for statistical studies 
\citep[e.g.][]{vagnetti10},
or simply as a convenient source of UV/optical photometry for small
samples of objects \citep[e.g. ][]{jin12} or individual sources
\citep[e.g. ][]{smith09}.

\section*{Acknowledgments}
\label{sec:acknowledgments}

This research has made use of the following archives: the USNOFS Image
and Catalogue Archive operated by the United States Naval Observatory,
Flagstaff Station; the Two Micron All Sky Survey, which is a joint
project of the University of Massachusetts and the Infrared Processing
and Analysis Center/California Institute of Technology, funded by the
National Aeronautics and Space Administration and the National Science
Foundation. We thank Tom Dwelly for supplying the Subaru B band 
image and sourcelist for the 13$^{H}$ deep field.

\bibliographystyle{mn2e}

\appendix

\vspace{3mm}
\noindent
{\bf APPENDIX A: TESTING AND VALIDATION OF THE OMDETECT SOURCE DETECTION 
ALGORITHM}
\vspace{1mm}

\begin{figure}
\includegraphics[width=65mm, angle=270]{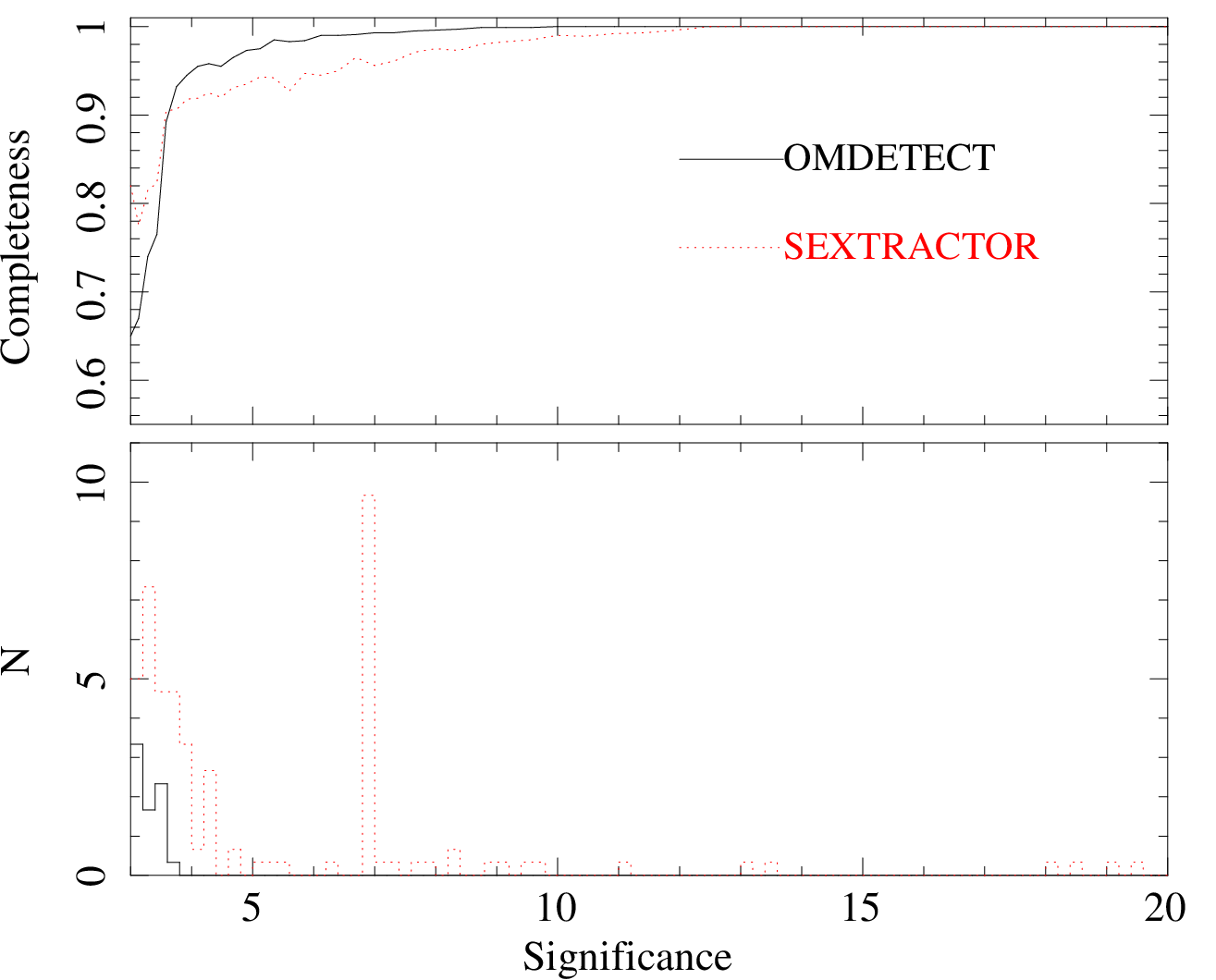}

\noindent
{\bf Figure A1. }
{The top panel shows the completeness, defined by (number of matched test
sources)/(number of test sources), of {\sc omdetect} and 
{\sc sextractor} in trials with simulated XMM-OM images as a function 
of source significance. The lower panel shows the number of spurious 
sources detected per 1000 fields as a function of source significance, from 
running {\sc omdetect} and {\sc sextractor} on 3000 simulated source-free 
fields.} 
\end{figure}

\noindent
{\bf A.1 Tests of the source detection algorithm on simulated images}
\vspace{1mm}

\noindent
The performance of the {\sc omdetect} source detection algorithm was
tested extensively on simulated data during the development cycle
leading up to the construction of the XMM-SUSS. To benchmark its
performance, 100 simulated images were produced with random background
levels between 0 and 200 counts per pixel and sources were placed at
random positions on the image.  Sources were either point-like, in
which case the source was constructed using the XMM-OM empirical
point-spread function for a given OM filter, or extended, in which
case the source had a 2-d Gaussian profile with a random position
angle. The brightness distribution of the sources was generated
randomly to simulate a true image. Each image had Poissonian noise
added.

The input test sources were matched with those detected by {\sc
  omdetect} to form a list of the properties of matched and unmatched
test sources. The output sources were then checked against the input
sources to ensure that the source positions, count rates, extension
flags, extended source widths and orientations are correctly recovered
by {\sc omdetect}. The simulation process was also used to check the
detection completeness, defined by (number of matched test
sources)/(number of test sources). The top panel of Fig. A1 shows the
completeness of {\sc omdetect} derived from the simulations as a
function of source significance. Comparison of the input and recovered
source lists indicate that the source detection algorithm is $> 95$
percent complete at a signal to noise level of $5$, and 100 percent
complete at a signal to noise ratio of $10$. As a measure of
performance, the results were compared to those obtained by running
{\sc sextractor} \citep{bertin96}, with parameters tuned for the
XMM-OM images, on the same simulated images. As can be seen in
Fig. A1, the completeness of {\sc omdetect} compares favourably with
that of {\sc sextractor}.

Tests of the source detection were also carried out on source-free
simulated images to assess the level of spurious sources. The results
of applying {\sc omdetect} and {\sc sextractor} to 3000 simulated
source-free images are shown in the lower panel of Fig. A1. Again,
{\sc omdetect} compares favourably with {\sc sextractor}.
\vspace{3mm}

\noindent
{\bf A.2 Tests of the source detection algorithm on XMM-OM images}
\vspace{1mm}

\begin{figure}
\includegraphics[width=85mm]{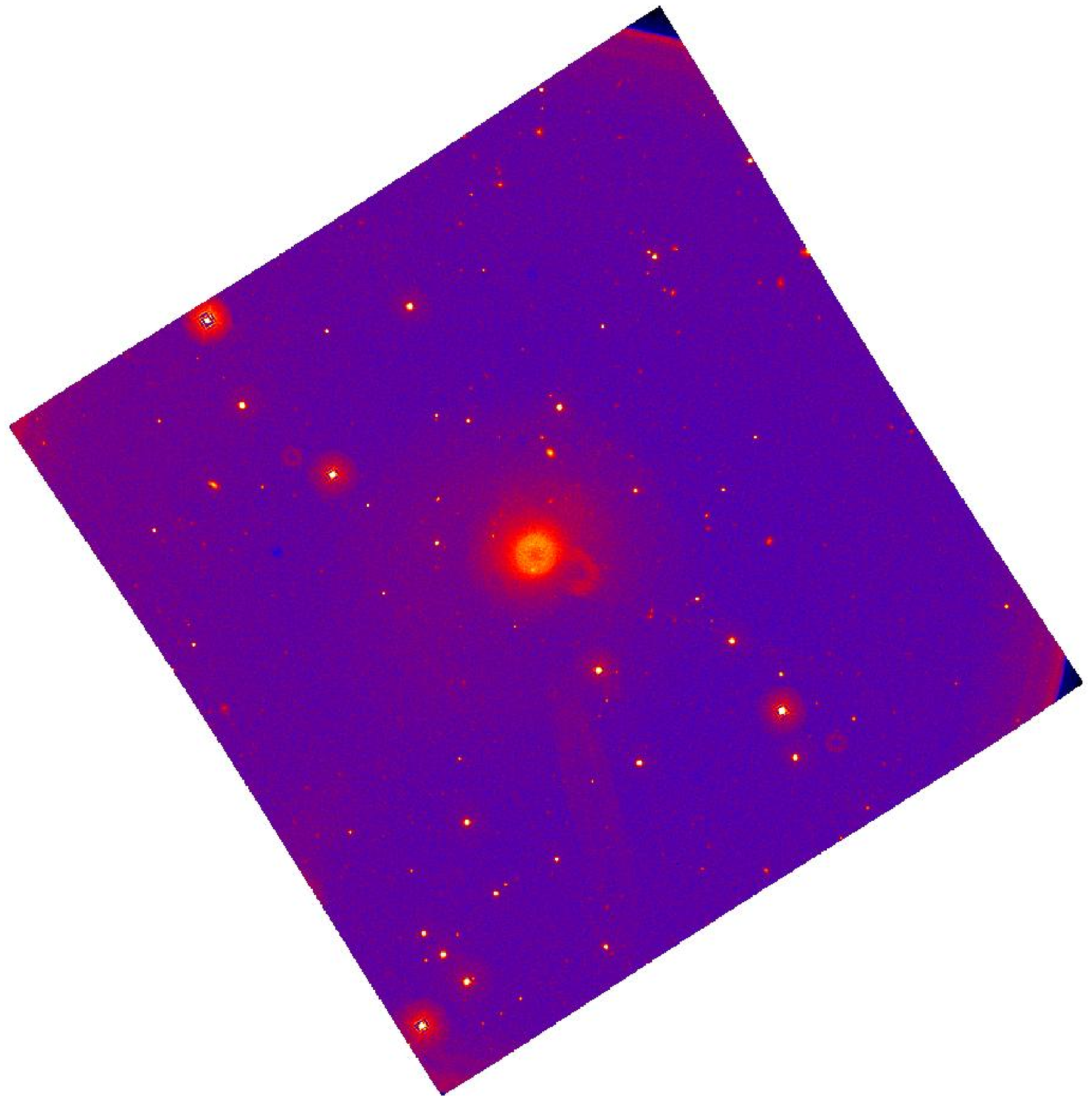}
\includegraphics[width=85mm]{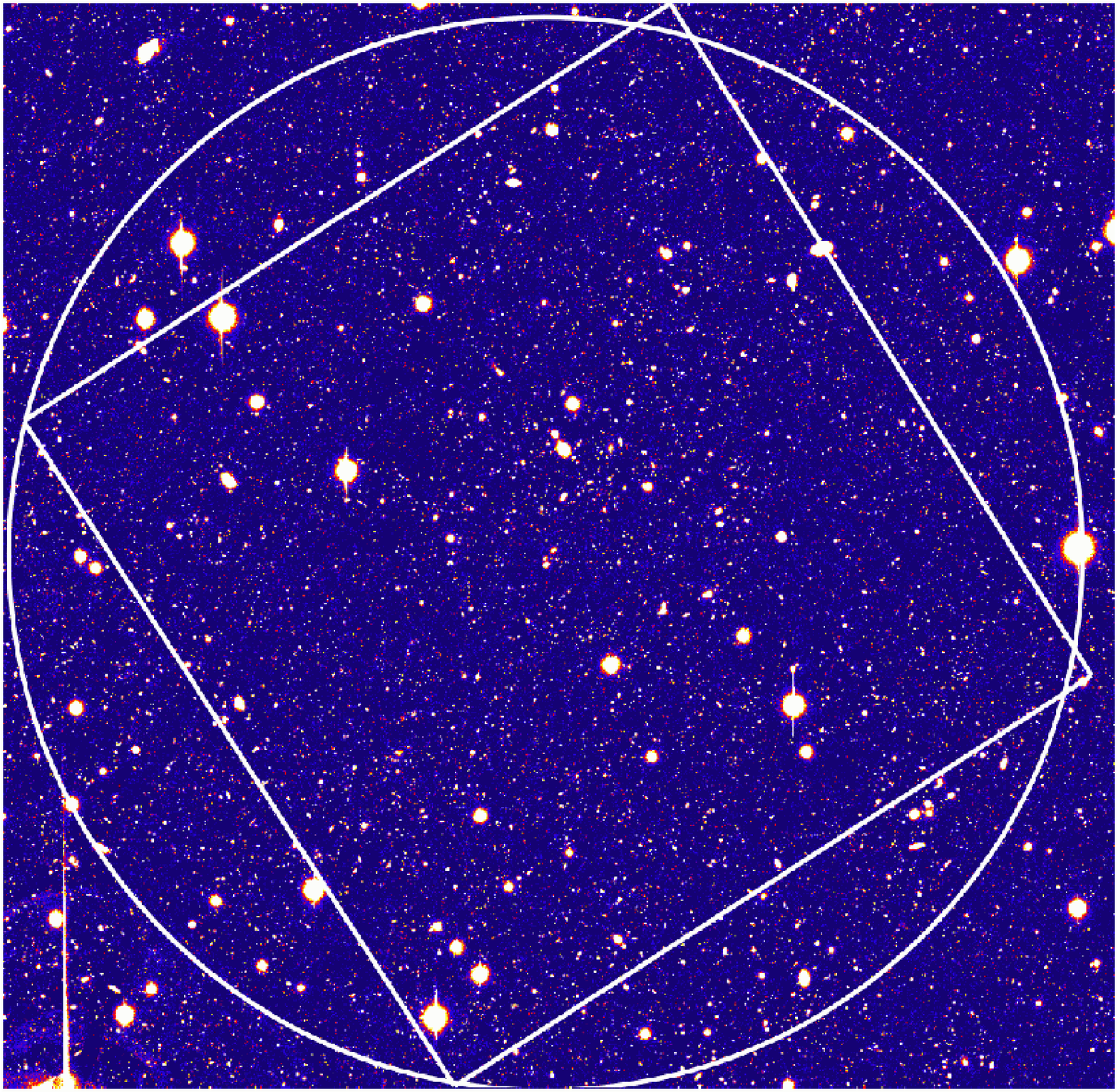}

\noindent 
{\bf Figure A2. } 
{The 13-hour {\it XMM-Newton}/{\it Chandra} deep field. The top panel
  contains the XMM-OM B-filter image. The bright, circular region at
  the centre of the image is the central enhanced region referred to
  in Section~\ref{subsubsec:scattered}.  The bottom panel contains the
  corresponding Subaru B-filter image with the OM field of view
  (diamond) overlayed. The XMM-OM is not sensitive to sources outside
  of the overlayed circle, hence XMM-OM and Subaru source detection
  characteristics are based only on sources within the overlapping
  region of the circle and diamond.  }
\end{figure}

Prior to the construction of the XMM-SUSS, {\sc omdetect} and {\sc
  sextractor} were run on a large number of OM images, and the
fidelity of the two source detection algorithms were visually examined
by overlaying the detected source regions on the images and looking
for missed, spurious and misclassified sources.  In agreement with the
simulations, {\sc omdetect} misses few sources, and those that are missed
nearly always have low signal-to-noise ratios ($<$4) and are usually
close to much brighter sources. In comparison, {\sc sextractor}
generated significant numbers of spurious sources in scattered light
features and coincidence-loss-induced modulo-8 distortion around
bright sources, and in images with very low backgrounds.\footnote{It
  should be noted that {\sc sextractor} was not designed to cope with
  any of these image characteristics.}

In order to provide a more quantitative validation of the 
fidelity of the {\sc omdetect} source
detection algorithm against an independent reference set, its
performance on a 5 ksec B-filter exposure of the field centered on RA
= 13$^{\mbox{h}}$ 34$^{\mbox{m}}$ 40$^{\mbox{s}}$.19, Dec =
$+$37$^\circ$ 54$\prime$ 58$\prime\prime$.9 (J2000) was compared to a
deeper, independent source list. This particular field is the 13$^{H}$
{\it XMM-Newton}/{\it Chandra} Deep Field and the independent source
list was obtained using SExtractor2.5.0 \citep{bertin96} 
from a Johnson B exposure
obtained by the 8-m Subaru telescope \citep[][Dwelly et~al., in
prep]{seymour08}.
The XMM-OM and Subaru images are shown in
Fig.~A2.  The point spread function of the
Subaru image has a full-width half-maximum of 0.87 arcsec and reaches
to much fainter magnitudes than the XMM-OM image.  The Subaru image
covers a larger sky area than the XMM-OM image, so the Subaru source
list was screened to remove objects that are outside the
XMM-OM field-of-view. Sources were matched between the two catalogues 
using the simple criterion of nearest source within 3 arcseconds.

\begin{figure*}
\includegraphics[width=165mm]{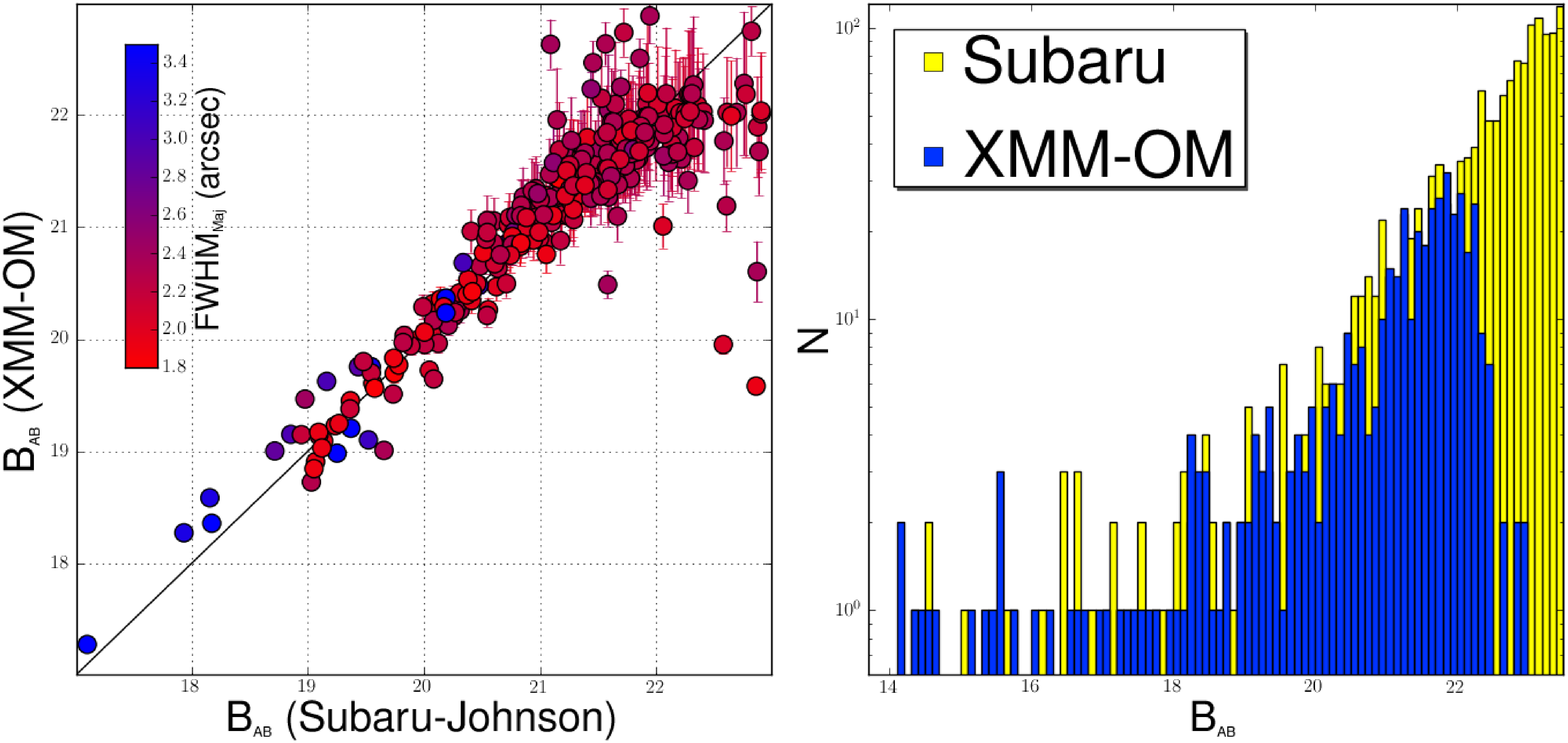}

\begin{minipage}[l]{2.1\columnwidth}
\noindent 
{\bf Figure A3. } 
{The left panel compares the B magnitudes of XMM-OM sources
  with their B Subaru counterparts. The Subaru filter is Johnson B,
  but the XMM-OM magnitudes have not been transformed to Johnson B
  because the accompanying XMM-OM data in U and V 
  do not go sufficiently deep to facilitate this for the faintest sources.
  Sources have been matched between the two catalogues
  using the simple criteria of nearest source within 3 arcsec. Colour
  coding is a function of source extent on the sky, point sources are
  red, extended sources are blue. Error bars correspond to 1$\sigma$
  confidence for the XMM-OM magnitudes.  The right hand panel provides
  a histogram of XMM-OM and Subaru source magnitudes.}
\end{minipage}
\end{figure*}

The left hand panel of Fig.~A3 compares the B
magnitudes of the XMM-OM sources with their nearest counterparts in
the Subaru list, while the right hand panel shows the magnitude
distributions of the XMM-OM and Subaru source lists.  Stars with
B$_{AB}<19$ mag are not used in the photometry comparison because they are
saturated in the Subaru image. The sources show a linear relation
between Subaru and XMM-OM magnitudes; outliers at faint magnitudes
come primarily from mis-identifications arising from the simplistic
cross-matching rather than photometric variability.
The Subaru and XMM-OM magnitude distributions indicate good detection
consistency between the two instruments. All the Subaru objects with
$B_{AB}<19$ are detected by {\sc omdetect} in the XMM-OM image,
although with different magnitudes because of the saturation in the
Subaru image. Furthermore, although there appears to be a small
deficit of XMM-OM sources between $20.5 < B_{AB} < 20.9$, all of these
Subaru objects are detected by {\sc omdetect}, with slightly
different magnitudes. The XMM-OM sourcelist is highly complete to
$B_{AB}=21.9$, corresponding to a signal to noise ratio of 4.5 for
compact sources.  Comparison with the Subaru image indicates that the
source detection process works well, and corroborates the high
completeness levels obtained in the simulations described at the
beginning of this section.

\vspace{3mm}
\noindent
{\bf APPENDIX B: CRITERIA USED FOR QUALITY FLAGS 1,2,3 AND 4}
\vspace{1mm}

\noindent
Here we describe the criteria used in the XMM-SUSS pipeline to
identify and flag readout streaks (flag 1), smoke rings (flag 2), diffraction
spikes (flag 3) and coincidence-loss distorted sources (flag 4), 
as described in Section \ref{subsec:flags}.

\vspace{3mm}
\noindent
{\bf B.1 Flag 1: readout streak}
\vspace{1mm}

\noindent
Any point-source
with  a raw  count-rate  $\geq  70$  \cntss\ or which is 
surrounded by coincidence-loss-induced  mod-8 pattern  (flag  bit 4)  
is  considered a  potential
source of  a readout  streak. Sources
situated within a threshold distance of the potential readout streak
are flagged. The threshold distance is 8 arcsec unless the read out
streak is produced by a source with a count rate $> 220$ \cntss, in
which case the threshold distance is set to 9.5 arcsec.  
The source responsible for a readout streak may not
be  located  on the  image because it may fall outside the data 
collection window.   A further  test  is  performed in  these
cases. The counts  in a raw image are described  by the array $C_{ij}$
where $i$  is the column number and  $j$ is the row  number.  Based on
data  screening, our  criteria  for a  column  containing a  potential
readout  streak   is  provided  by   the  six  conditions   $C_{ij}  >
C_{\left(i-k\right)j}$ where $k  = -3$, $-2$, $-1$, $1$,  $2$, $3$. If
each crtierion is met individually over $>$60 per cent of the pixel rows then
any source situated within 19 rows of the column are flagged as
occurring over a potential readout streak.
\vspace{3mm}

\noindent
{\bf B.2 Flag 2: smoke ring}
\vspace{1mm}

\noindent
Sources containing raw count rates $\geq 60$ \cntss\ or surrounded by
a coincidence-loss-induced modulo-8 pattern (flag bit 4) are
considered generators of smoke-rings.  A smoke ring is located
approximately along a radial line from the field centre through the
star which generates it. The smoke ring is displaced towards the field
edge by an amount which is non-linear with off-axis angle. The
location of each ring in the raw image pixels ($i_{sr}$,$j_{sr}$) is
predicted using the following parameterization:
\begin{equation}
i_{sr} = \left(A_1 + i_c\right) + A_2 x + A_3 y + A_4 x^2 + A_5 y^2 + A_6 x y
\end{equation}
\begin{equation}
j_{sr} = \left(B_1 + j_c\right) + B_2 x + B_3 y + B_4 x^2 + B_5 y^2 + B_6 x y
\end{equation}
where ($i_{\hbox{c}}$,$j_{\hbox{c}}$) = (1024.5,1024.5) is the center
of the FOV, and ($x$,$y$) = ($i_* - i_{\hbox{c}}$,$j_* -
j_{\hbox{c}}$).  The coefficients $A_{1...6}$ and $B_{1...6}$ were
determined from a least squares fit to a random sample of 100 smoke
rings, and are given in Table B1.

\begin{table}
{\bf Table B1.} Coefficients for smoke ring locations.
\begin{tabular}{l@{\hspace{15mm}}l@{\hspace{15mm}}l}
\hline
$n$&$A_{n}$&$B_{n}$\\
\hline
&&\\
1&9.6950                &$-4.8165$\\
2&1.2052                &$5.0884\times 10^{-4}$\\
3&1.1027$\times 10^{-3}$&1.2071\\
4&2.9117$\times 10^{-6}$&$-2.8810\times 10^{-7}$\\
5&1.8790$\times 10^{-6}$&$1.5490\times 10^{-6}$\\
6&4.2437$\times 10^{-7}$&$-1.0783\times 10^{-6}$\\
\hline
\end{tabular}
\end{table}

\vspace{3mm}

\noindent
{\bf B.3 Flag 3: diffraction spike}
\vspace{1mm}

\noindent
All catalogue objects with either raw count-rates $\geq 70$ \cntss\ or
flagged as being surrounded by a coincidence-loss-induced modulo-8
pattern (flag bit 4) have been tested for diffraction spikes.  We
compare source number densities \nbox\ and \nspike where \nbox\ is the
density of objects with raw count rates $< 5$ \cntss\ within a box
$70\times70$ arcsec$^2$ centred on the bright source, and \nspike\ is
the average density of objects with raw count rates $< 5$ \cntss\
within four rectangular regions oriented radially away from the bright
source in directions 45$^\circ$, 135$^\circ$, 225$^\circ$ and
315$^\circ$ of dimensions $(10 \times 52$) arcsec$^2$, starting 5
arcsec from the position of the bright source. These four rectangular
regions correspond to the locations of the putative diffraction
spikes. If $\nspike > 3 \nbox$, all detections $< 5$ arcsec from the
diffraction spikes are flagged.  Note that this algorithm will 
inevitably be less effective in crowded fields. \vspace{3mm} 

\noindent
{\bf B.4 Flag 4: bright source surrounded by coincidence-loss-induced 
modulo-8 pattern}
\vspace{1mm}

\noindent
If the brightest pixel within a 4 arcsec x 4 arcsec box centered on a
detection has a count rate $> 10$ counts~s$^{-1}$ then the object is
flagged as a bright source with surrounding mod-8 structure. If the
brightest pixel has a count rate which is $< 10$ counts~s$^{-1}$, but
$> 0.5$ counts~s$^{-1}$ then the source undergoes a further test: a
pattern search algorithm is performed on the regions $5-15$ arcsecond
to the left, right, above and below the source in raw image
coordinates.  The pattern search checks for a 10-arcsec-long column to
the left or right of the source, or a 10-arcsec-long row above or
below the source, which has all of its pixels lower in value than all
of the pixels in one of the surrounding columns or rows. If such a
pattern is identified in two or more of the four sides of the source,
then the source is flagged.

\vspace{3mm}
\noindent
{\bf APPENDIX C: COLUMNS IN THE XMM-SUSS SOURCE TABLE}
\vspace{1mm}

\noindent
The XMM-SUSS fits table contains two extensions. The first extension
provides the source catalogue, and the second extension lists the
XMM-Newton observations from which the catalogue was constructed.  The
column headings for the XMM-SUSS source catalogue table are listed in
Table C1, and we now provide a brief description of the
columns. Filter-specfic entries are set to `NULL' for filters in which
the source is not detected.  Column 1 gives the source name in
International Astronomical Union format. Columns 2 and 3 refer to the
{\em XMM-Newton} observation in which the source was found; column 2
is the corresponding row number in the second extension of the fits
file (which lists the {\em XMM-Newton} observations) and column 3
gives the 10 digit observation identification number (OBSID) of the
observation. Column 4 gives the source number within the merged source
list deriving from that observation 
(this number is only unique to the source within the relevant 
{\em XMM-Newton} observation, not within the XMM-SUSS as a whole). 
Column 5 is a unique reference
number for the source within the XMM-SUSS. Columns 6--11 gives the
distance in arcsec of the source to the nearest celestial source
in the catalogue which is detected in the named filter. Note that as 
the catalogue only contains sources which are detected in the UV, the nearest 
source in any filter must (also) have been detected in at least 1 UV band. 
Columns 12--15 give the right ascention and
declination of the source in decimal and hexagesimal formats. Column
16 gives the 1-$\sigma$ position uncertainty in arcsec. Columns 17 and
18 give the Galactic longitude and latitude in degrees. Column 19
gives the number of different {\em XMM-Newton} observations in which
the source has been detected. 
  Since a source is given a separate row for each {\em XMM-Newton}
  observation in which it is detected, column 19 also corresponds to
  the number of rows in which the source appears in the XMM-SUSS.
Columns 20--25 give the number of
exposures within the {\em XMM-Newton} observation in which the source
is detected in each filter. Columns 26--31 give the detection
significance of the source in each filter. Columns 32--43 give the
count rates and uncertainties in each filter. Columns 44--55 give, for
each filter, the flux density and uncertainty at the effective 
wavelength (Table \ref{tab:filters}) of the filter.
Columns 56--67 give the AB magnitudes and errors, and columns 68--79
the Vega-based magnitudes and errors.  Columns 80--91 give for each
filter the $\chi^{2}/\nu$ value for a constant source countrate
and the maximum deviation from the median
countrate in terms of $\sigma$, for sources detected multiple times 
through the same filter, in the same {\em XMM-Newton} observation 
(see Section \ref{subsec:TimeSampling}).  
Columns 92--109 give, for each
filter, the sizes of the major and minor axes of the source in arcsec
and the position angle of the major axis.  Columns 110--115 give the
quality flags as integer numbers (derived from the sum of the flag bit
values) for the different filters and columns 116--121 give the
quality flags as strings of logical values corresponding to the flag
bits (T and F when the flag is set or not set respectively). Columns
122--127 state whether the source appears to be pointlike (value 0) or
extended (value 1) in each of the filters.

\begin{table*}
{\bf Table C1.} Column names in the XMM-SUSS table.\\
\begin{tabular}{llll}
&&&\\
Column number & Column name & \vline\ \ \ Column number & Column name \\
              &             & \vline\ \ \               &             \\            
1   &  IAUNAME                    & \vline\ \ \ 65  &  B$\_$AB$\_$MAG$\_$ERR          \\
2   &  N$\_$SUMMARY               & \vline\ \ \ 66  &  V$\_$AB$\_$MAG              \\
3   &  OBSID                      & \vline\ \ \ 67  &  V$\_$AB$\_$MAG$\_$ERR          \\
4   &  SRCID                      & \vline\ \ \ 68  &  UVW2$\_$VEGA$\_$MAG         \\
5   &  SRCNUM                     & \vline\ \ \ 69  &  UVW2$\_$VEGA$\_$MAG$\_$ERR     \\
6   &  UVW2$\_$SRCDIST            & \vline\ \ \ 70  &  UVM2$\_$VEGA$\_$MAG         \\
7   &  UVM2$\_$SRCDIST            & \vline\ \ \ 71  &  UVM2$\_$VEGA$\_$MAG$\_$ERR     \\
8   &  UVW1$\_$SRCDIST            & \vline\ \ \ 72  &  UVW1$\_$VEGA$\_$MAG         \\
9   &  U$\_$SRCDIST               & \vline\ \ \ 73  &  UVW1$\_$VEGA$\_$MAG$\_$ERR     \\
10  &  B$\_$SRCDIST               & \vline\ \ \ 74  &  U$\_$VEGA$\_$MAG            \\
11  &  V$\_$SRCDIST               & \vline\ \ \ 75  &  U$\_$VEGA$\_$MAG$\_$ERR        \\
12  &  RA                         & \vline\ \ \ 76  &  B$\_$VEGA$\_$MAG            \\
13  &  DEC                        & \vline\ \ \ 77  &  B$\_$VEGA$\_$MAG$\_$ERR        \\
14  &  RA$\_$HMS                  & \vline\ \ \ 78  &  V$\_$VEGA$\_$MAG            \\
15  &  DEC$\_$DMS                 & \vline\ \ \ 79  &  V$\_$VEGA$\_$MAG$\_$ERR        \\
16  &  POSERR                     & \vline\ \ \ 80  &  UVW2$\_$CHI2             \\
17  &  LII                        & \vline\ \ \ 81  &  UVW2$\_$MAXDEV           \\
18  &  BII                        & \vline\ \ \ 82  &  UVM2$\_$CHI2             \\
19  &  N$\_$OBSID                 & \vline\ \ \ 83  &  UVM2$\_$MAXDEV           \\
20  &  N$\_$UVW2$\_$EXP           & \vline\ \ \ 84  &  UVW1$\_$CHI2             \\
21  &  N$\_$UVM2$\_$EXP           & \vline\ \ \ 85  &  UVW1$\_$MAXDEV           \\
22  &  N$\_$UVW1$\_$EXP           & \vline\ \ \ 86  &  U$\_$CHI2                \\
23  &  N$\_$U$\_$EXP              & \vline\ \ \ 87  &  U$\_$MAXDEV              \\
24  &  N$\_$B$\_$EXP              & \vline\ \ \ 88  &  B$\_$CHI2                \\
25  &  N$\_$V$\_$EXP              & \vline\ \ \ 89  &  B$\_$MAXDEV              \\
26  &  UVW2$\_$SIGNIF             & \vline\ \ \ 90  &  V$\_$CHI2                \\
27  &  UVM2$\_$SIGNIF             & \vline\ \ \ 91  &  V$\_$MAXDEV              \\
28  &  UVW1$\_$SIGNIF             & \vline\ \ \ 92  &  UVW2$\_$MAJOR$\_$AXIS       \\
29  &  U$\_$SIGNIF                & \vline\ \ \ 93  &  UVM2$\_$MAJOR$\_$AXIS       \\
30  &  B$\_$SIGNIF                & \vline\ \ \ 94  &  UVW1$\_$MAJOR$\_$AXIS       \\
31  &  V$\_$SIGNIF                & \vline\ \ \ 95  &  U$\_$MAJOR$\_$AXIS          \\
32  &  UVW2$\_$RATE               & \vline\ \ \ 96  &  B$\_$MAJOR$\_$AXIS          \\
33  &  UVW2$\_$RATE$\_$ERR        & \vline\ \ \ 97  &  V$\_$MAJOR$\_$AXIS          \\
34  &  UVM2$\_$RATE               & \vline\ \ \ 98  &  UVW2$\_$MINOR$\_$AXIS       \\
35  &  UVM2$\_$RATE$\_$ERR        & \vline\ \ \ 99  &  UVM2$\_$MINOR$\_$AXIS       \\
36  &  UVW1$\_$RATE               & \vline\ \ \ 100 &  UVW1$\_$MINOR$\_$AXIS       \\
37  &  UVW1$\_$RATE$\_$ERR        & \vline\ \ \ 101 &  U$\_$MINOR$\_$AXIS          \\
38  &  U$\_$RATE                  & \vline\ \ \ 102 &  B$\_$MINOR$\_$AXIS          \\
39  &  U$\_$RATE$\_$ERR           & \vline\ \ \ 103 &  V$\_$MINOR$\_$AXIS          \\
40  &  B$\_$RATE                  & \vline\ \ \ 104 &  UVW2$\_$POSANG           \\
41  &  B$\_$RATE$\_$ERR           & \vline\ \ \ 105 &  UVM2$\_$POSANG           \\
42  &  V$\_$RATE                  & \vline\ \ \ 106 &  UVW1$\_$POSANG           \\
43  &  V$\_$RATE$\_$ERR           & \vline\ \ \ 107 &  U$\_$POSANG              \\
44  &  UVW2$\_$AB$\_$FLUX         & \vline\ \ \ 108 &  B$\_$POSANG              \\
45  &  UVW2$\_$AB$\_$FLUX$\_$ERR  & \vline\ \ \ 109 &  V$\_$POSANG              \\
46  &  UVM2$\_$AB$\_$FLUX         & \vline\ \ \ 110 &  UVW2$\_$QUALITY$\_$FLAG     \\
47  &  UVM2$\_$AB$\_$FLUX$\_$ERR  & \vline\ \ \ 111 &  UVM2$\_$QUALITY$\_$FLAG     \\
48  &  UVW1$\_$AB$\_$FLUX         & \vline\ \ \ 112 &  UVW1$\_$QUALITY$\_$FLAG     \\
49  &  UVW1$\_$AB$\_$FLUX$\_$ERR  & \vline\ \ \ 113 &  U$\_$QUALITY$\_$FLAG        \\
50  &  U$\_$AB$\_$FLUX            & \vline\ \ \ 114 &  B$\_$QUALITY$\_$FLAG        \\
51  &  U$\_$AB$\_$FLUX$\_$ERR     & \vline\ \ \ 115 &  V$\_$QUALITY$\_$FLAG        \\
52  &  B$\_$AB$\_$FLUX            & \vline\ \ \ 116 &  UVW2$\_$QUALITY$\_$FLAG$\_$ST  \\
53  &  B$\_$AB$\_$FLUX$\_$ERR     & \vline\ \ \ 117 &  UVM2$\_$QUALITY$\_$FLAG$\_$ST  \\
54  &  V$\_$AB$\_$FLUX            & \vline\ \ \ 118 &  UVW1$\_$QUALITY$\_$FLAG$\_$ST  \\
55  &  V$\_$AB$\_$FLUX$\_$ERR     & \vline\ \ \ 119 &  U$\_$QUALITY$\_$FLAG$\_$ST     \\
56  &  UVW2$\_$AB$\_$MAG          & \vline\ \ \ 120 &  B$\_$QUALITY$\_$FLAG$\_$ST     \\
57  &  UVW2$\_$AB$\_$MAG$\_$ERR   & \vline\ \ \ 121 &  V$\_$QUALITY$\_$FLAG$\_$ST     \\
58  &  UVM2$\_$AB$\_$MAG          & \vline\ \ \ 122 &  UVW2$\_$EXTENDED$\_$FLAG    \\
59  &  UVM2$\_$AB$\_$MAG$\_$ERR   & \vline\ \ \ 123 &  UVM2$\_$EXTENDED$\_$FLAG    \\
60  &  UVW1$\_$AB$\_$MAG          & \vline\ \ \ 124 &  UVW1$\_$EXTENDED$\_$FLAG    \\
61  &  UVW1$\_$AB$\_$MAG$\_$ERR   & \vline\ \ \ 125 &  U$\_$EXTENDED$\_$FLAG       \\
62  &  U$\_$AB$\_$MAG             & \vline\ \ \ 126 &  B$\_$EXTENDED$\_$FLAG       \\
63  &  U$\_$AB$\_$MAG$\_$ERR      & \vline\ \ \ 127 &  V$\_$EXTENDED$\_$FLAG       \\
64  &  B$\_$AB$\_$MAG             & \vline\ \ \     &                        \\
\end{tabular}
\end{table*}

\label{lastpage}

\end{document}